# CHARACTERISTIC FEATURES OF THE RESONANT TRIDENT PROCESS IN THE FIELD OF A STRONG MONOCHROMATIC ELECTROMAGNETIC WAVE

**S.P. Roshchupkin[1,*], M.V. Shakhov[1]**

[1] Higher School of Fundamental Physical Research, Peter the Great St. Petersburg Polytechnic University, Saint-Petersburg, Russian Federation

\* Correspondence author; E-mail: serg9rsp@gmail.com.

**Abstract:** The characteristic features of the resonant trident process (Oleinik resonances) have been theoretically studied in a wide range of frequencies and intensities of a circularly polarized strong electromagnetic wave. The resonant trident process is defined by two characteristic quantum energies: the characteristic energy of the nonlinear Compton effect and the characteristic energy of the nonlinear Breit-Wheeler process. These characteristic energies depend significantly on the frequency and intensity of the wave, as well as on the angle between the momenta of the initial electrons and the electromagnetic wave. The resonant trident process is effective when the energy of the initial electrons is greater than or on the order of magnitude of the corresponding characteristic energies. It is shown that quantum entanglement of final particles takes place in this resonant process. An important aspect of the resonant trident process is the equality of the energies of the electron and positron pairs. Analytical expressions for the differential rates of the resonant trident process on the energy of final particles are obtained. The corresponding analytical expressions for full rates have also been obtained. It is shown that the rate data of the resonant trident process in the field of optical and X-ray wave frequencies significantly exceed the corresponding rate of the non-resonant trident process. Results obtained can be used in experiments at leading laser centers, as well as to explain QED processes in strong X-ray fields near neutron stars and magnetars.

**Keywords:** Oleinik resonances, the resonant trident process, nonlinear Compton effect, nonlinear Breit-Wheeler process, ultra-relativistic electrons and positrons, quantum entanglement.

**1. Introduction.**

Development of high-intensity lasers raised interest in QED in external fields (see articles [1-18]). Among these experiments LUXE collaboration should be highlighted as it is set to give essential data for electron-laser and gamma-laser interactions in strong electromagnetic fields (see articles [1,5-7]). The main points of interest in QED in external fields are the first-order processes: the Nonlinear Compton Effect (NCE) (see articles [19-30]) and Nonlinear Breit-Wheeler Process (NBWP) (see articles [31-47]). But in recent years the second-order processes by the fine-structure constant raised in popularity among QED scientists with the studies of the Trident Process: the electron and electron-positron pair production in electron-laser collisions (see articles [48-57]). It should be noted that second-order processes can be described in two different ways: the resonant (Oleinik resonances) [58-65] and non-resonant way [50-57]. We emphasize that the resonant processes are essential for describing cosmology problems (see articles [66-72]) where high wave frequencies and high initial particle energies are present. The resonant studies can be applied to many more problems in external fields like nuclei- and vacuum-laser interactions (see articles [73-87]).

It should be noted that the non-resonant case also includes the article [56], which studied positron energy distribution in the factorized trident process. In this article, the trident process is proposed to be studied as a two-stage process of electron-positron pair generation. On the first stage as a result of a NCE high-energy photon appears. On the second stage the NBWP creates electron-positron pairs. It is important to note that here a high-energy photon is observable. On the other hand, the trident process can proceed in a resonant way when the intermediate gamma quantum becomes real. In this case, the initial resonant trident process (RTP) also effectively splits into two first-order processes according to the fine structure constant: the nonlinear Compton effect and the nonlinear Breit-Wheeler process. It is important to note the qualitative difference between the resonant trident process and the factorized trident process, due to the fact that in the first case, the intermediate real gamma quantum is unobservable, in the quantum mechanical sense, and in the second case, the gamma quantum is observable. This is what leads to different kinematics and significantly different rates of these processes.

In this article, we will focus on developing the results obtained in our previous paper [60] on the resonant trident process. We obtain qualitatively new results that significantly expand the understanding of the resonant kinematics of RTP, and for the first time we obtain differential RTP rates for the energy of final particles, as well as the total RTP rate for various initial electron energies, wave frequencies and intensities. Further, we will be using relativistic system of units: $\hbar = c = 1$.

Let us choose the 4-potential of the external electromagnetic field in the form of a plane circularly polarized monochromatic wave propagating along the z axis:

$$A(\varphi) = \frac{F}{\omega} \cdot \left( e_x \cos\varphi + \delta e_y \sin\varphi \right), \quad \varphi = (kx) = \omega(t-z), \quad \delta = \pm 1. \qquad (1)$$

Here $F$ is electric field strength, $\omega$ - frequency of the wave, $k = (\omega, \mathbf{k})$ is a 4-wave vector, $e_x, e_y$ - 4-vectors of wave polarization that have the following properties: $e_x = (0, \mathbf{e_x})$, $e_y = (0, \mathbf{e_y})$, $e_x e_y = 0$, $(e_x)^2 = (e_y)^2 = -1$. The main parameter of the discussed problem is the classical relativistic-invariant parameter

$$\eta = \frac{eF}{m\omega}. \qquad (2)$$

which is numerically equal to ratio of the field work at the wavelength to rest energy of electron ($e$ and $m$ are charge and mass of electron). In this problem we have relativistic-invariant quantum parameters of the NCE $(\varepsilon_{iC})$ and the NBWP $(\varepsilon_{iBW})$ [60-64]:

$$\varepsilon_{iC} = \frac{2(kp_i)}{m_*^2}, \quad \varepsilon_{iBW} = \frac{(kp_i)}{2m_*^2}. \tag{3}$$

Here $p_i = (E_i, \mathbf{p}_i)$ is the initial electron 4-momentum and $m_*$ is the effective electron mass in the field of a circularly polarized wave

$$m_* = m\sqrt{1+\eta^2}. \tag{4}$$

In the discussed problem relativistic-invariant parameters, related to the final particles – electron and electron-positron pair, are defined as:

$$\varepsilon_{jC} = \frac{2(kp_j)}{m_*^2}, \quad \varepsilon_{jBW} = \frac{(kp_j)}{2m_*^2}, \quad j = f, \pm. \tag{5}$$

Further, we will be discussing ultra-relativistic case of the initial electron energy $(E_i)$ and the energies of the final particles: electron $(E_f)$ and electron-positron pair $(E_\mp)$.

$$E_j \gg m, \quad j = i, f, \mp. \tag{6}$$

In ultra-relativistic case (6) final particles scatter in a narrow cone along the momentum of the initial electron. At the same time, we will assume that this narrow cone of particles lies far from the direction of propagation of the external electromagnetic wave (otherwise, the resonances disappear [60-65]). Therefore, the angles between the momenta of particles: the initial electrons $(\mathbf{p}_i)$, the final electrons $(\mathbf{p}_f)$ and the electrons (positrons) of the pair $(\mathbf{p}_\mp)$, must satisfy the following relations:

$$\theta_f \equiv \angle(\mathbf{p_i}, \mathbf{p}_f) \ll 1, \quad \theta_\pm \equiv \angle(\mathbf{p}_+, \mathbf{p}_-) \ll 1, \tag{7}$$

$$\theta_i \equiv \angle(\mathbf{p_i}, \mathbf{k}) \sim 1, \quad \angle(\mathbf{p}_j, \mathbf{k}) \approx \theta_i, \quad j = f, \mp. \tag{8}$$

We will also assume that the classical parameter $\eta$ is limited from above by the following condition [60-64]:

$$\eta \ll \min\left(\frac{E_f}{m}, \frac{E_\pm}{m}\right). \tag{9}$$

Because of this, further consideration of the resonance process will be valid for sufficiently high wave intensities. However, the intensity of these fields should be less than the critical Schwinger field $F_* \approx 1.3 \cdot 10^{16}$ V/cm.

In kinematic conditions (6) – (8) quantum parameters of the NCE and the NBWP (3), as well as quantum parameter (5) have a simple physical meaning. They are equal to the ratio of the corresponding energy of the initial or final particles to the characteristic quantum energy of the NCE or the NBWP:

$$\varepsilon_{jC} = \frac{E_j}{\omega_C}, \quad \varepsilon_{jBW} = \frac{E_j}{\omega_{BW}}, \quad j = i, f, \pm. \tag{10}$$

Here $\omega_C$ and $\omega_{BW}$ are the characteristic quantum energies of the NCE and the NBWP correspondingly [60-64]:

$$\omega_C = \frac{m^2(1+\eta^2)}{4\omega \sin^2(\theta_i/2)}, \quad \omega_{BW} = 4\omega_C. \tag{11}$$

From here it follows that characteristic quantum energies (11) are inversely proportional to the frequency $(\omega)$ and proportional to the intensity of the external electromagnetic wave $(I \sim \eta^2)$. They also depend on angle between initial electron momentum and external wave-vector. Note that at $\theta_i \to 0$ characteristic energies $\omega_C \to \infty, \omega_{BW} \to \infty$. Therefore, quantum parameters $\varepsilon_{iC} \to 0, \varepsilon_{iBW} \to 0$ and Oleinik resonances disappear [60-65]. In the articles [1, 5, 56] discussed experiment conditions (for example, for LUXE-0, $\omega = 1.55$ eV, $\eta = [0.1; 5]$, $\theta_i = 17.2°$, $E_i = 16.5$ GeV) even for weak fields $(\eta = 0.1)$) characteristic energy of the NCE (11) $\omega_C \approx 2$ TeV and quantum parameter $\varepsilon_{iC} \approx 0.83 \times 10^{-2}$. In this case, the resonant conditions (see relation (54)) threshold number of photons of the wave for the NBWP will be very large $(n_{th} \approx 5 \times 10^4)$. Therefore, the resonant trident process will be suppressed. Because of that the trident process described in this article has non-resonant character.

Further, we will assume that the characteristic energy of the NCE significantly exceeds the rest energy of the electron.

$$\frac{\omega_C}{m} \sim \frac{m}{\omega}(1+\eta^2) \gg 1. \tag{12}$$

Condition (12) imposes a restriction on the frequencies of the external wave. The frequencies should not be close to the rest energy of the electron. It is important to note that in the resonant case, the inverse of the NCE quantum parameter determines the optimal number of photons of the wave (at which the probability of the process has the maximum value) in the NCE (at the first vertex) [61]:

$$n_{iC} = \lceil \varepsilon_{iC}^{-1} \rceil. \tag{13}$$

At the same time, the inverse of the quantum parameter of the NBWP determines the necessary condition for the minimum number of photons of the wave in the NBWP (at the second vertex) [60-64]:

$$n_{iBW} = \lceil \varepsilon_{iBW}^{-1} \rceil. \tag{14}$$

As will be shown below, condition (14) is a necessary but not a sufficient condition in the resonance process (see sections 3.2 and 3.3 of this article). We also note that for ultra-relativistic particle energies and small outgoing angles, it is common (see, for example, articles [56, 60-64]) ultra-relativistic parameters and energies normalized to the energy of the initial particles are used. Thus, in [60], ultra-relativistic parameters and energies normalized to the energy of the initial electrons were used for the outgoing angles of the final electron and the electron-positron pair, as well as their energies:

$$\delta_j = \frac{E_i \theta_j}{m_*}, \quad x_j = \frac{E_j}{E_i}, \quad j = f, \pm. \tag{15}$$

At the same time, depending on the resonant differential probability of the outgoing angles, there is no obvious dependence on the energy of the initial electrons. At the same time, the resonance process has characteristic quantum energies of the NCE (at the first vertex) and the NBWP (at the second vertex), which are determined only by the parameters of the electromagnetic wave and the rest energy of the electron (11). Because of this, it is advisable for the RTP to normalize the energies, as well as the outgoing angles of particles, with the characteristic energies of the NCE and the NBWP. So, for the ultra-relativistic parameters (15), we obtain:

$$\delta_f \to \varepsilon_{iC}\delta_{fC}, \quad \delta_{fC} = \frac{\omega_C \theta_f}{m_*}; \quad \delta_{\pm} \to \varepsilon_{iBW}\delta_{\pm BW}, \quad \delta_{\pm BW} = \frac{\omega_{BW}\theta_{\pm}}{m_*}. \tag{16}$$

With this in mind, the dependence on the energy of the initial electrons in the RTP rate will be shown explicitly through quantum parameters $\varepsilon_{iC}$ and $\varepsilon_{iBW}$. Therefore, in this article energies and outgoing angles of the final particles will be normalized by characteristic energies of the NCE and the NBWP.

## 2. Amplitude of the resonant trident process.

The amplitude of the process can be written as (see [60]):

$$S_{if} = \sum_{n_1=1}^{+\infty}\sum_{n_2=1}^{+\infty} S_{n_1 n_2}, \tag{17}$$

where partial amplitude $S_{n_1 n_2}$ corresponds to absorption of $n_1$ (in the first vertex) and $n_2$ (in the second vertex) – photons of the external wave.

$$S_{n_1 n_2} = \frac{16\pi^5 e^2 e^{-id_0}}{\sqrt{\tilde{E}_i \tilde{E}_f \tilde{E}_- \tilde{E}_+}} \left\{ \left[ \bar{u}_{p_f} F^\mu_{-n_1}(\tilde{p}_f,\tilde{p}_i) u_{p_i} \right] \frac{1}{q^2} \left[ \bar{u}_{p_-} F_{-n_2,\mu}(\tilde{p}_-,-\tilde{p}_+) v_{p_+} \right] - (\tilde{p}_f \Leftrightarrow \tilde{p}_-) \right\} \times$$
$$\delta^{(4)}\left[ \tilde{p}_+ + \tilde{p}_- + \tilde{p}_f - \tilde{p}_i - (n_1+n_2)k \right]. \tag{18}$$

Here 4-momentum of the intermediate photon is defined as
$$q = \tilde{p}_+ + \tilde{p}_- - n_2 k. \tag{19}$$

In expression (18) $u_{p_i}$ – Dirac's bispinor of the initial electron, $\bar{u}_{p_f}, \bar{u}_{p_-}, v_{p_+}$ – Dirac's bispinors of the final electron and electron-positron wave, $d_0$ – phase independent of summation indices, function $F^\mu_{-n_1}(\tilde{p}_f,\tilde{p}_i)$ has the following form:

$$F^\mu_{-n_1}(\tilde{p}_f,\tilde{p}_i) = a^\mu(\tilde{p}_f,\tilde{p}_i) L_{-n_1} + b^\mu_-(\tilde{p}_f,\tilde{p}_i) L_{-n_1-1} + b^\mu_+(\tilde{p}_f,\tilde{p}_i) L_{-n_1+1}. \tag{20}$$

Here matrices $a^\mu$ and $b^\mu_\pm$ are defined as:

$$a^\mu(\tilde{p}_f,\tilde{p}_i) = \gamma^\mu + \frac{\eta^2 m^2}{2(k\tilde{p}_f)(k\tilde{p}_i)} \hat{k} k^\mu, \tag{21}$$

$$b^\mu_\pm(\tilde{p}_f,\tilde{p}_i) = \frac{\eta m}{4}\left[ \frac{1}{(k\tilde{p}_f)} \hat{\varepsilon}_\pm \hat{k} \gamma^\mu + \frac{1}{(k\tilde{p}_i)} \gamma^\mu \hat{k} \hat{\varepsilon}_\pm \right], \tag{22}$$

$$\varepsilon_\pm = e_x \pm i e_y. \tag{23}$$

In expression (20) special function $L_{-n_1}$ [60-64], that in the case of circular polarization of the external wave can be expressed with Bessel functions of integer order

$$L_{-n_1} = \exp(in_1 \chi_{\tilde{p}_f \tilde{p}_i}) J_{-n_1}(\gamma_{\tilde{p}_f \tilde{p}_i}), \qquad (24)$$

where indicated

$$\gamma_{\tilde{p}_f \tilde{p}_i} = \eta m_* \sqrt{-Q^2_{\tilde{p}_f \tilde{p}_i}}, \qquad (25)$$

$$\tan \chi_{\tilde{p}_f \tilde{p}_i} = \delta \frac{(Q_{\tilde{p}_f \tilde{p}_i} e_y)}{(Q_{\tilde{p}_f \tilde{p}_i} e_x)}, \quad Q_{\tilde{p}_f \tilde{p}_i} = \frac{\tilde{p}_f}{(k\tilde{p}_f)} - \frac{\tilde{p}_i}{(k\tilde{p}_i)}. \qquad (26)$$

In the expressions above $m_*$ is the effective mass of the electron in the plane wave field (4), $\tilde{p}_j = (\tilde{E}_j, \tilde{\mathbf{p}}_j)$ are 4-quasimomenta of electrons and positrons:

$$\tilde{p}_j = p_j + \eta^2 \frac{m^2}{2(kp_j)} k, \quad i = f, \pm. \qquad (27)$$

It should be noted, that functions $F_{-n_2,\mu}(\tilde{p}_-, -\tilde{p}_+)$ in expression (18) are obtained from (20) – (26) by substituting: $\tilde{p}_i \to -\tilde{p}_+$, $\tilde{p}_f \to -\tilde{p}_-$, $n_1 \to n_2$, and lowering $\mu$ indices.

## 3. Resonant kinematics of the resonant trident process.

In case of resonance, intermediate gamma-quantum becomes real (Oleinik resonances [58-65]). As a result, initial second-order process by the fine-structure constant in the wave field effectively splits into two first-order processes by the fine-structure constant: NCE (in the first vertex) [19-31] and NBWP (in the second vertex) [32-47]. The diagram of the process in the resonant case is illustrated on figure 1.

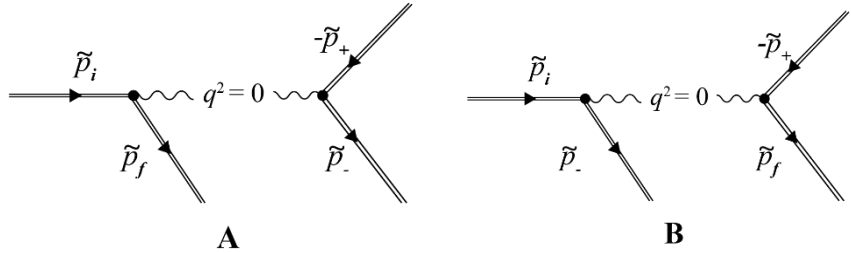

**Figure 1**. Feynman diagram of the resonant trident process: channels A and B.

$$q^2 = 0. \qquad (28)$$

The following expressions will be considered for the resonant channel A. Let us define the conservation law of the 4-momentum in the first and second vertices of the Feynman diagram in resonant case (28) (see fig.1A):

$$\tilde{p}_i + n_1 k = q + \tilde{p}_f, \qquad (29)$$

$$q + n_2 k = \tilde{p}_+ + \tilde{p}_-. \qquad (30)$$

Considering that $\tilde{p}_{i,f}^2 = \tilde{p}_{\pm}^2 = m_*^2$ and $q^2 = k^2 = 0$, relations (29) and (30) are valid if $n_1 \geq 1$ and $n_2 \geq 1$. Note that relation (29) is the conservation law of the 4-momentum for NCE, and expression (30) is the conservation law of the 4-momentum for NBWP.

### 3.1. Kinematics of the resonant trident process on the lower branches of the final particle energies

In the first vertex the NCE takes place. Considering the conservation law in the first vertex (29), as well as the resonance condition (28), we obtain relativistic-invariant expression for the 4-momenta of the initial and final electrons in the electromagnetic wave field:

$$\frac{2(p_i p_f)}{m_*^2} + \frac{\eta^2}{(1+\eta^2)}\left(\frac{\varepsilon_{iC}}{\varepsilon_{fC}} + \frac{\varepsilon_{fC}}{\varepsilon_{iC}}\right) = 2 + n_1(\varepsilon_{iC} - \varepsilon_{fC}). \tag{31}$$

Here relativistic-invariant parameters $\varepsilon_{iC}$ and $\varepsilon_{fC}$ have the form (3) and (5). In kinematic conditions (6) – (8), (10) we can obtain the quadratic equation for the resonant energy of the final electron in the first vertex from expression (31). The solution for this equation will be two possible branches of the electron energy (the lower branch with «minus» sign before the square root and the upper branch with «plus» sign before the square root, see relation (32)). It should be mentioned, that electron energy on the upper branch does not contain physical cense at the zero outgoing angle of the electron (see [60]), whilst the lower branch is defined on the whole range of possible outgoing angles. The final electron energy can be presented in units of characteristic energy of the NCE:

$$\varepsilon_{fC(d)} = \frac{E_{f(d)}}{\omega_C} = \frac{2 + n_1 \varepsilon_{iC}\left(1 - \sqrt{1 - 4\delta_{fC(d)}^2/n_1^2}\right)}{2\left(n_1 + \varepsilon_{iC}^{-1} + \varepsilon_{iC}\delta_{fC(d)}^2\right)}. \tag{32}$$

Here $\delta_{fC(d)}^2$ is the ultra-relativistic parameter of the electron (16), that defines the square of its outgoing angle at the lower branch of energy. Depending on the number of absorbed photons in the first vertex, this parameter changes in the following interval:

$$0 \leq \delta_{fC(d)}^2 \leq \left(\frac{n_1}{2}\right)^2, \quad \delta_{fC(d)}^2 = \frac{\omega_C^2 \theta_f^2}{m_*^2}. \tag{33}$$

From expression (32) it follows that the energy of the final electron (in units of characteristic energy of the NCE) on the lower branch is defined by the number of absorbed photons of the wave in the first vertex $n_1$, the quantum parameter of the NCE $\varepsilon_{iC}$ (10), (11) and the ultra-relativistic parameter of the electron $\delta_{fC(d)}^2$ (33). The minimum and the maximum of the resonant electron energy on the lower branch is obtained from relation (32) at zero and maximum outgoing angle:

$$E_{f(d)}^{\min} = \frac{E_i}{1 + n_1 \varepsilon_{iC}}, \tag{34}$$

$$E_{f(d)}^{\max} = \frac{E_i}{1 + n_1 \varepsilon_{iC}/2}, \quad n_1 \geq 1. \tag{35}$$

Considering relations (34), (35), expression for the final electron energy on the lower branch (32) can be rewritten in a more pleasing for analysis form:

$$\varepsilon_{fC(d)} = \left[\frac{1}{\varepsilon_{iC}} + \frac{n_1}{\rho_{f(d)}}\right]^{-1}, \qquad (36)$$

where function $\rho_{f(d)}$ is defined as

$$\rho_{f(d)} = \frac{2 + n_1\varepsilon_{iC}\left(1 - \sqrt{1 - 4\delta_{fC(d)}^2/n_1^2}\right)}{1 + 2\varepsilon_{iC}\delta_{fC(d)}^2/n_1 + \sqrt{1 - 4\delta_{fC(d)}^2/n_1^2}}. \qquad (37)$$

With the change of the ultra-relativistic parameter $\delta_{fC(d)}^2$ from zero to the maximum (33), function $\rho_{f(d)}$ changes from 1 to 2 (see expressions (34), (36)):

$$1 \leq \rho_{f(d)} \leq 2. \qquad (38)$$

In the second vertex in the resonant case the NBWP takes place. Considering the conservation law in the second vertex (30), as well as the resonance condition (28), we obtain relativistic-invariant expression for the 4-momenta of the electron-positron pair in the electromagnetic wave field:

$$2\left[1 + \frac{(p_+ p_-)}{m_*^2}\right] + \frac{\eta^2}{(1+\eta^2)}\left(\frac{\varepsilon_{+BW}}{\varepsilon_{-BW}} + \frac{\varepsilon_{-BW}}{\varepsilon_{+BW}}\right) = 4n_2\left(\varepsilon_{+BW} + \varepsilon_{-BW}\right). \qquad (39)$$

Here relativistic-invariant parameters $\varepsilon_{\pm BW}$ have the form (5). Expression (39) is invariant to the symmetrical substitution of the 4-momenta of the electron-positron pair $(p_+ \leftrightarrow p_-)$. Therefore, we can consider energies of the pair to be equal $(E_+ = E_-)$. In kinematic conditions (6) – (8), (10) we can obtain the quadratic equation for the resonant energy of the electron-positron pair in the second vertex from expression (39). The solution for this equation will be two possible branches of the electron-positron pair energy (the lower branch with «minus» sign before the square root and the upper branch with «plus» sign before the square root, see relation (40)). It should be mentioned, that electron-positron pair energy on the upper branch does not contain physical cense at the zero outgoing angle (see [60]), whilst the lower branch is defined on the whole range of possible outgoing angles. The electron-positron pair energy can be presented in units of characteristic energy of the NBWP:

$$\varepsilon_{\pm BW(d)} = \frac{E_{\pm(d)}}{\omega_{BW}} = \frac{\rho_{\pm(d)}}{2n_2}, \quad \rho_{\pm(d)} = \frac{2n_2^2}{\delta_{\pm BW(d)}^2}\left(1 - \sqrt{1 - \frac{\delta_{\pm BW(d)}^2}{n_2^2}}\right). \qquad (40)$$

Here $\delta_{\pm BW(d)}^2$ is the ultra-relativistic parameter of the pair, that defines the square of its outgoing angle at the lower branch of energy. Depending on the number of absorbed photons in the second vertex, this parameter changes in the following interval:

$$0 \leq \delta_{\pm BW(d)}^2 \leq n_2^2, \quad \delta_{\pm BW(d)}^2 = \frac{\omega_{BW}^2 \theta_\pm^2}{4m_*^2}. \qquad (41)$$

From expression (40) it follows that the positron (electron) energy of the pair (in units of characteristic energy of the NBWP) on the lower branch is defined by the number of absorbed photons of the wave in the second vertex $n_2$ and the ultra-relativistic parameter of the pair $\delta^2_{\pm BW(d)}$. Note that the change of the ultra-relativistic parameter of the pair $\delta^2_{\pm BW(d)}$ from zero to the maximum corresponds to the change of the function $\rho_{\pm(d)}$ (40) from 1 to 2:

$$1 \leq \rho_{\pm(d)} \leq 2. \quad (42)$$

The minimum and the maximum of the positron (electron) energy on the lower branch is obtained from relation (40) at zero and the maximum outgoing angle:

$$E^{\min}_{\pm(d)} = \frac{E_i}{2n_2 \varepsilon_{iBW}} = \frac{\omega_{BW}}{2n_2}, \quad E^{\max}_{\pm(d)} = \frac{E_i}{n_2 \varepsilon_{iBW}} = \frac{\omega_{BW}}{n_2}. \quad (43)$$

Considering that the maximum energy of the pair must be at least less than the energy of the initial electron, we obtain the following restriction on the minimum number of absorbed photons in the second vertex:

$$n_2 \geq \left\lceil \frac{2}{\varepsilon_{iBW}} \right\rceil. \quad (44)$$

Note that condition (44) is necessary but not sufficient (see sections 3.2 and 3.3 of this article). Expression for $\rho_{\pm(d)}$ (40) can be rewritten in a more pleasing for further analysis of the threshold number of photons of the wave absorbed in the second vertex for the NBWP. Let us set parametric change of the ultra-relativistic parameter of the pair $\delta^2_{\pm BW(d)}$ in the interval (41):

$$\delta^2_{\pm BW(d)} = \tau n_2^2, \quad 0 \leq \tau \leq 1. \quad (45)$$

The expressions for $\rho_{\pm(d)}$ function and the positron (electron) of the pair energy on the lower branches (40) are defined in a parametric form:

$$\rho_{\pm(d)} \to \rho_\tau = \frac{2}{\tau}\left(1 - \sqrt{1-\tau}\right), \quad (46)$$

$$E_{\pm(d)} = \frac{\rho_\tau}{2n_2} \omega_{BW}. \quad (47)$$

Therefore, resonant energies of the electron and the electron-positron pair have two possible branches of solutions. However, only lower branches of the final particles energies have physical cense in the whole range of possible outgoing angles (see relations (32) and (40)). Because of the general energy conservation law of the RTP, every energy of the electron (pair) on the lower branch has a corresponding energy of the pair (electron) on the upper branch. Therefore, the conservation law of the RTP defines corresponding energies on the upper branches. Further, we will be distinguishing two possible combinations of resonant energies and angles of final particles. Therefore, we have two cases of quantum entanglement of the final particle's states. Let us consider these two cases.

### 3. 2 Quantum entanglement at electron energy (outgoing angle) detection

In this case the resonant energy of the final electron is defined on its lower branch (32), (36), (37). The energy of the electron-positron pair on the upper branch is defined from the general energy conservation law of the RTP:

$$2E_{\pm(u)} \approx E_i - E_{f(d)}. \tag{48}$$

Diving every term in relation (48) by $\omega_C$ we obtain:

$$\varepsilon_{\pm C(u)} = \frac{E_{\pm(u)}}{\omega_C} = 4\frac{E_{\pm(u)}}{\omega_{BW}}, \quad \delta^2_{\pm C(u)} = \frac{\omega_C^2 \theta_\pm^2}{m_*^2}. \tag{49}$$

Substituting the upper branch for the energy of the electron-positron pair (see relation (40) with a «plus» before the square root) and lower branch for the final electron energy (32) into the general energy conservation law (48), considering (49), we obtain the following expression:

$$\sqrt{1-4\delta^2_{\pm C(u)}/n_2^2} = \frac{1}{2n_2}\delta^2_{\pm C(u)}\left(\varepsilon_{iC} - \varepsilon_{fC(d)}\right) - 1 > 0. \tag{50}$$

From this we get the ulta-relativistic parameter of the pair $\delta^2_{\pm C(u)}$, that is uniquely defined by the energy of the final electron on the lower branch, number of absorbed photons of the wave in the second vertex and the quantum parameter of the NCE.

$$\delta^2_{\pm C(u)} = \kappa_{f(d)}\left(n_2 - \kappa_{f(d)}\right). \tag{51}$$

Here function $\kappa_{f(d)}$ has the following form:

$$\kappa_{f(d)} = \frac{4}{\left(\varepsilon_{iC} - \varepsilon_{fC(d)}\right)}. \tag{52}$$

It is important to note that function $\kappa_{f(d)}$ (52) can be defined by $\rho_{f(d)}$ (37) if expressions (36), (37) are used for the electron energy.

$$\kappa_{f(d)} = \frac{4}{\varepsilon_{iC}}\left(1 + \frac{\rho_{f(d)}}{n_1 \varepsilon_{iC}}\right). \tag{53}$$

Expressions (51), (53) define relation between outgoing angle of the pair and the outgoing angle of the electron. Substituting expression (51) in the right part of relation (50), we get the following restriction on the number of photons of the wave absorbed in the second vertex in dependence on outgoing angle of the electron:

$$n_2 \geq n_{th}, \quad n_{th} = \lceil 2\kappa_{f(d)} \rceil. \tag{54}$$

Here $n_{th}$ is the threshold value of the number of photons of the wave absorbed in the second vertex (in the NBWP). Relation (51) allows us to define the interval of change for the ultra-relativistic parameter $\delta^2_{\pm C(u)}$ for the upper branch of the energy of the electron-positron pair in dependence on the outgoing angle of the final electron:

$$\delta^2_{\pm C(\min)} \leq \delta^2_{\pm C(u)} \leq \delta^2_{\pm C(\max)}, \tag{55}$$

where indicated

$$\delta^2_{\pm C(\min)} = \kappa_{f(\min)}\left(n_2 - \kappa_{f(\min)}\right), \quad \kappa_{f(\min)} = \frac{4}{\varepsilon_{iC}}\left(1 + \frac{1}{n_1\varepsilon_{iC}}\right), \tag{56}$$

$$\delta^2_{\pm C(\max)} = \kappa_{f(\max)}\left(n_2 - \kappa_{f(\max)}\right), \quad \kappa_{f(\max)} = \frac{4}{\varepsilon_{iC}}\left(1 + \frac{2}{n_1\varepsilon_{iC}}\right). \tag{57}$$

The interval of change in the threshold value of the number of photons of the wave in the second vertex $n_{th}$ (54) in dependence on the outgoing angle of the final electron is defined by the expression:

$$n_{th(\min)} \leq n_{th} \leq n_{th(\max)}. \tag{58}$$

Here indicated:

$$n_{th(\min)} = \lceil 2\kappa_{f(\min)} \rceil, \quad n_{th(\max)} = \lceil 2\kappa_{f(\max)} \rceil. \tag{59}$$

From relation (58) we can see that in the whole range of the outgoing angles of the electron-positron pair on the upper branch its energy (55) the number of photons of the wave absorbed in the second vertex must satisfy the following condition:

$$n_2 \geq n_{th(\max)}. \tag{60}$$

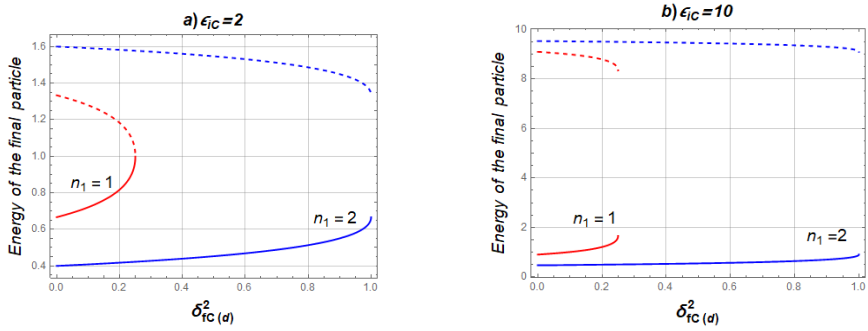

**Figure 2**. Dependence of the energy of the final electron $\varepsilon_{fC(d)}$ on the lower branch (32) (solid curves) and the electron-positron pair $2\varepsilon_{\pm C(u)}$ on the upper branch (48) (hatched curves) on the ultra-relativistic parameter of the electron $\delta^2_{fC(d)}$ with fixed number of the photons of the wave absorbed in the first vertex $(n_1 = 1, 2)$. Fig.2a and fig.2b correspond the energy of the initial electron $\varepsilon_{iC} = 2$ and $\varepsilon_{iC} = 10$ respectively.

It can be seen from Figure 2 that with an increase in the number of absorbed photons of the wave, the range of outgoing angles of the final electron increases (see relation (33)). The energy of the final electron decreases, and the energy of the electron (positron) of the pair increases. At the same time, an increase in the energy of the initial electron also leads to a decrease in the energy of the final electron and an increase in the energy of the pair.

We can determine the minimum energy of the initial electron, at which the threshold number of the photons of the wave absorbed in the second vertex is equal to one over the whole range of the electron outgoing angles.

$$\varepsilon_{iC} \geq \varepsilon_{n_1(\min)} = 4\left(1 + \sqrt{1 + \frac{1}{n_1}}\right), \quad (n_1 \geq 1, n_2 \geq 1). \tag{61}$$

From expression (61) it follows that at $n_1 = 1$ and $\varepsilon_{iC} \geq \varepsilon_{1(\min)} \approx 9.68$ process in the second vertex appears with absorption of one photon of the wave. In this case, as it will be shown further, the resonant rate of the process will be the biggest. If $\varepsilon_{iC} \ll 1$, then the threshold number of photons of the wave (54), (59) has the following order magnitude: $n_{th} \sim \lceil n_1^{-1} \varepsilon_{iC}^{-2} \rceil \gg 1$. In this case the RTP rate will be suppressed. When the electron scatters at the zero angle $\left(\delta_{fC(d)}^2 = 0\right)$ the outgoing angle of the pair will be minimal ($\delta_{\pm C(u)}^2 = \delta_{\pm C(\min)}^2$, see expression (56)). In these conditions the number of the photons of the wave absorbed in the second vertex will be defined by a «milder» condition.

$$n_2 \geq n_{th(\min)}. \tag{62}$$

From this, we obtain a condition for the possible energies of the initial electron $\varepsilon_{iC}$, for which the minimum number of photons in the second vertex is equal to one:

$$\varepsilon_{iC} \geq \varepsilon'_{n_1(\min)} = 4\left(1 + \sqrt{1 + \frac{1}{2n_1}}\right) \quad (n_1 \geq 1, n_2 \geq 1). \tag{63}$$

Because of this, when scattering final particles at the minimum angles for $n_1 = 1$ and $\varepsilon_{iC} \geq \varepsilon'_{1(\min)} \approx 8.92$, the process of absorption of photons of the wave in the second vertex begins with a single photon. Thus, the energy of the positron (electron) of the pair on the upper branch, because of the general law conservation of energy, is determined by the energy of the electron on the lower branch (36), (37) or the outgoing angle of the electron:

$$\varepsilon_{\pm C} = \frac{1}{2}\left(\varepsilon_{iC} - \varepsilon_{fC}\right) = \frac{\varepsilon_{iC}}{2}\left(1 + \frac{\rho_f}{n_1 \varepsilon_{iC}}\right)^{-1}. \tag{64}$$

In this case, the maximum and minimum energies of an electron-positron pair take the form:

$$E_{\pm(u)}^{\max} = \frac{E_i}{2}\left(1 + \frac{1}{n_1 \varepsilon_{iC}}\right)^{-1}, \quad E_{\pm(u)}^{\min} = \frac{E_i}{2}\left(1 + \frac{2}{n_1 \varepsilon_{iC}}\right)^{-1}. \tag{65}$$

Let us consider an experimental situation where the energy of the initial electrons significantly exceeds the characteristic energy of the NCE. In this case, the NCE quantum parameter satisfies the condition:

$$\varepsilon_{iC} \gg 1. \tag{66}$$

Then the energies of the final particles (36), (64) will take the form:

$$E_{f(d)} \approx \frac{\rho_f}{n_1}\omega_{iC} \ll E_i, \quad E_{\pm(u)} \approx \frac{1}{2}\left(E_i - \frac{\rho_f}{n_1}\omega_{iC}\right) \approx \frac{E_i}{2}. \tag{67}$$

In this case the energy of the final electron is much smaller than the energy of the initial electron and has the order of magnitude of the characteristic energy of the NCE. In this case, the positron and electron energies of the pair are equal and close to half the energy of the initial electrons. It is important to emphasize that in this case, the ultra-relativistic parameter of the pair $\delta^2_{\pm C(u)}$ (51)-(53) is close to the value of $\delta^2_{\pm C(*)}$ regardless of the outgoing angle of the electron:

$$\delta^2_{\pm C(u)} \approx \delta^2_{\pm C(*)} = \frac{4}{\varepsilon_{iC}}\left(n_2 - \frac{4}{\varepsilon_{iC}}\right), \quad (n_2 \geq 1). \tag{68}$$

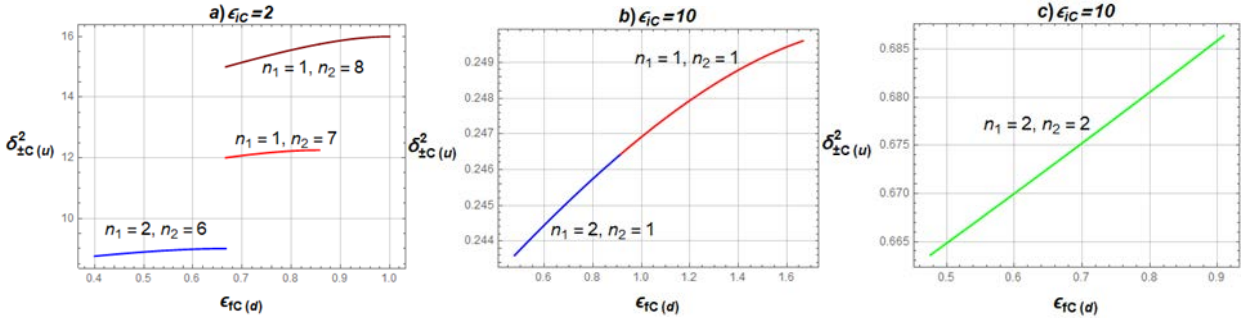

**Figure 3**. Dependence of the ultra-relativistic parameter of the pair $\delta^2_{\pm C(u)}$ on the upper branch (51) on the electron energy on the lower branch $\varepsilon_{fC(d)}$ (34) for different numbers of the photons of the wave in the first $(n_1 \geq 1)$ and second $(n_2 \geq n_{th})$ (54) vertices. Рис.3а отвечает энергии начального электрона $\varepsilon_{iC} = 2$. Figure 3a corresponds to the energy of the initial electron $\varepsilon_{iC} = 2$. Figure 3b and Figure 3c correspond to the number of absorbed photons of the wave $n_2 = 1$ $(n_1 = 1, 2)$ and $n_2 = 2$ $(n_1 = 2)$, respectively, for the initial electron energy $\varepsilon_{iC} = 10$.

Figure 3 shows the dependence of the square of the outgoing angle of the electron-positron pair on the upper branch of its energies (51) as a function of the energy of the final electron (34). Figure 3a corresponds to small energies of the initial electron $(\varepsilon_{iC} = 2)$, whereas the threshold number of the photons of the wave absorbed in the second vertex significantly depends on the energy of the final electron (the number of the photons of the wave absorbed at the first vertex). So, for $n_1 = 1$ we get $n_2 = 7, 8, \ldots$, and for $n_1 = 2$ we have $n_2 = 6, 7, \ldots$ (see relation (54)). Fig.3b and Fig.3c are plotted for large energies of the initial electron $(\varepsilon_{iC} = 10)$, when the number of the photons absorbed in the second vertex starts from one. Figure 3 shows that each possible interval of change in the final electron energy uniquely corresponds to a certain interval of change in the outgoing angle of the pair.

### 3.3 Quantum entanglement at electron energy (outgoing angle) detection.

In this case, the resonant energy of an electron-positron pair is determined by its lower branch (40), (45), (46). The electron energy on the upper branch is determined from the energy conservation law for the RTP:

$$E_{f(u)} = E_i - 2E_{\pm(d)}. \tag{69}$$

Dividing each term in the relation (69) by $\omega_{BW}$ and taking into account the equalities:

$$\varepsilon_{fBW} = \frac{E_{f(u)}}{\omega_{BW}} = \frac{1}{4}\frac{E_{f(u)}}{\omega_C}, \quad \delta^2_{fBW} = \frac{\omega^2_{BW}\theta^2_f}{4m^2_*} = 4\delta^2_{fC}. \tag{70}$$

Substituting into the energy conservation law (69) corresponding energies of the final particles (see the expression (40) and the relation (32), in which plus sign is put before the square root), we get the following relation:

$$\sqrt{1-\delta_{fBW(u)}^2/n_1^2} = 2\left(1+4\varepsilon_{iBW}n_1+4\varepsilon_{iBW}^2\delta_{fBW(u)}^2\right)\left(1-\frac{\rho_\tau}{n_2\varepsilon_{iBW}}\right) - 2\left(1+2n_1\varepsilon_{iBW}\right) > 0. \quad (71)$$

Here the function $\rho_\tau$ takes form (46). After simple transformations, we obtain the desired dependence of the outgoing angle of the final electron on the upper branch on the outgoing angle of the electron-positron pair on the lower branch:

$$\delta_{fBW(u)}^2 = 4\kappa_{\pm(d)}\left(n_1-\kappa_{\pm(d)}\right), \quad (72)$$

where the function $\kappa_{\pm(d)}$ is defined by the expression:

$$\kappa_{\pm(d)} = \frac{1}{4\varepsilon_{iBW}}\left(\frac{n_2\varepsilon_{iBW}}{\rho_\tau}-1\right)^{-1}. \quad (73)$$

Relation (72) allows us to determine the range of changes in the ultra-relativistic parameter $\delta_{fBW(u)}^2$ for the upper branch of the final electron energy:

$$\delta_{fBW(\min)}^2 \leq \delta_{fBW(u)}^2 \leq \delta_{fBW(\max)}^2. \quad (74)$$

where is indicated

$$\delta_{fBW(\min)}^2 = 4\kappa_{\pm(\min)}\left(n_1-\kappa_{\pm(\min)}\right), \quad \kappa_{\pm(\min)} = \frac{1}{4\varepsilon_{iBW}\left(n_2\varepsilon_{iBW}-1\right)}, \quad (75)$$

$$\delta_{fBW(\max)}^2 = 4\kappa_{\pm(\max)}\left(n_1-\kappa_{\pm(\max)}\right), \quad \kappa_{\pm(\max)} = \frac{1}{2\varepsilon_{iBW}\left(n_2\varepsilon_{iBW}-2\right)}. \quad (76)$$

In this case, the energy of the electron on the upper branch is completely determined by the energy of the electron-positron pair on the lower branch:

$$\varepsilon_{fBW(u)} = \frac{E_{f(u)}}{\omega_{BW}} = \varepsilon_{iBW}\left(1-\frac{\rho_\tau}{n_2\varepsilon_{iBW}}\right). \quad (77)$$

Hence, it is easy to obtain expressions for the maximum and minimum energy of the final electron:

$$E_{f(u)}^{\min} = E_i\left(1-\frac{2}{n_2\varepsilon_{iBW}}\right), \quad E_{f(u)}^{\max} = E_i\left(1-\frac{1}{n_2\varepsilon_{iBW}}\right). \quad (78)$$

The number of the photons of the wave absorbed in the second vertex $n_2$ in the relations (72) - (78) is determined by substituting the expressions (72), (45) into the right side of the relation (71). After simple calculations, we get the desired relation for $n_2$:

$$n_2 \geq \bar{n}_{th}. \quad (79)$$

Here $\bar{n}_{th}$ is the threshold number of photons of the wave absorbed in the second vertex, which is determined by the function $\rho_\tau$ (46):

$$\bar{n}_{th} = \lceil \zeta_\pm \rceil, \quad \zeta_\pm = \frac{\rho_\tau}{\varepsilon_{iBW}}\left(1+\frac{1}{2n_1\varepsilon_{iBW}}\right). \quad (80)$$

We emphasize that in the relation (80), the change of the parameter $\tau$ from zero to one determines the interval for change of the threshold number of photons of the wave $\bar{n}_{th}$, as well as the ultra-relativistic parameter $\delta^2_{\pm BW(d)}$ (45):

$$\bar{n}_{th(\min)} \leq \bar{n}_{th} \leq \bar{n}_{th(\max)}, \tag{81}$$

where

$$\bar{n}_{th(\min)} = \lceil \zeta_{\pm(\min)} \rceil, \quad \bar{n}_{th(\max)} = \lceil 2\zeta_{\pm(\min)} \rceil, \tag{82}$$

$$\zeta_{\pm(\min)} = \frac{1}{\varepsilon_{iBW}}\left(1 + \frac{1}{2n_1\varepsilon_{iBW}}\right). \tag{83}$$

Thus, the threshold number of photons of the wave absorbed in the second vertex at the minimum outgoing angles of the electron (59) and electron-positron pair (82) differ from each other. This is due to various features at minimal outgoing angles on the upper branches of the electron and pair energies. In this case, the threshold number of photons of the wave absorbed in the second vertex at the maximum outgoing angles of the electron (59) and electron-positron pair (82) equal to each other. Thus, when an electron-positron pair flies out at the zero angle $\left(\delta^2_{\pm BW(d)} = 0\right)$, the number of photons of the wave absorbed in the second vertex will be determined by a "softer" condition:

$$n_2 \geq \bar{n}_{th(\min)}. \tag{84}$$

From this, we obtain the condition for possible values of the energy of the initial electron $\varepsilon_{iBW}$, for which the minimum number of photons of the wave absorbed in the second vertex is equal to one:

$$\varepsilon_{iBW} \geq \bar{\varepsilon}_{n_1(\min)} = \frac{1}{2}\left(1 + \sqrt{1 + \frac{2}{n_1}}\right), \quad (n_1 \geq 1, n_2 \geq 1). \tag{85}$$

Thus, when scattering at the minimum angles of the final particles, the process of absorption of photons of the wave in the second vertex begins with a single photon for $n_1 = 1$ and $\varepsilon_{iBW} \geq \bar{\varepsilon}_{1(\min)} \approx 1.37$.

Now we consider the case when the energy of the initial electrons is significantly higher than the characteristic energy of the NBWP. In this case, the quantum parameter of the NBWP satisfies the condition:

$$\varepsilon_{iBW} \gg 1, \quad (\varepsilon_{iC} \gg 1). \tag{86}$$

Under the condition (86) resonant energies of the final particles (47), (77) take the form:

$$E_{\pm(d)} = \frac{\rho_\tau}{2n_2}\omega_{iBW} \ll E_i, \quad E_{f(u)} = E_i - \frac{\rho_\tau}{n_2}\omega_{iBW} \approx E_i. \tag{87}$$

In this case, the energy of the positron and electron of the pair is of the same order of magnitude of the characteristic energy of the NBWP and much smaller than the energy of the initial electron. On the other hand, the energy of the final electron is close to the energy of the initial electron. It is important to emphasize that in this case, the ultra-relativistic parameter of the electron on the upper branch of energy $\delta^2_{fBW(u)}$ (72) - (73) takes the following form:

$$\delta^2_{fBW(u)} \approx \frac{\rho_\tau}{\varepsilon^2_{iBW}}\left(\frac{n_1}{n_2}\right) \ll 1. \tag{88}$$

It shows that depending on the outgoing angle of the electron-positron pair (45), the outgoing angle of the electron becomes twice as big, while remaining small:

$$\delta^2_{fBW(*)} \leq \delta^2_{fBW(u)} \leq 2\delta^2_{fBW(*)}, \quad \delta^2_{fBW(*)} = \frac{1}{\varepsilon^2_{iBW}}\left(\frac{n_1}{n_2}\right) \ll 1. \tag{89}$$

To sum up the resonant kinematics, we can note that there are two different cases. The first case occurs, when the states of the final particles (their energy and outgoing angles) are mainly determined by the NCE, that occurs at the first vertex. The second case occurs, when the states of the final particles (their energy and outgoing angles) are mainly determined by the NBWP, that occurs at the second vertex. Thus, in the RTP, quantum entanglement of the sates of the final particles occurs, which is determined by two different first-order processes (the NCE or NBWP).

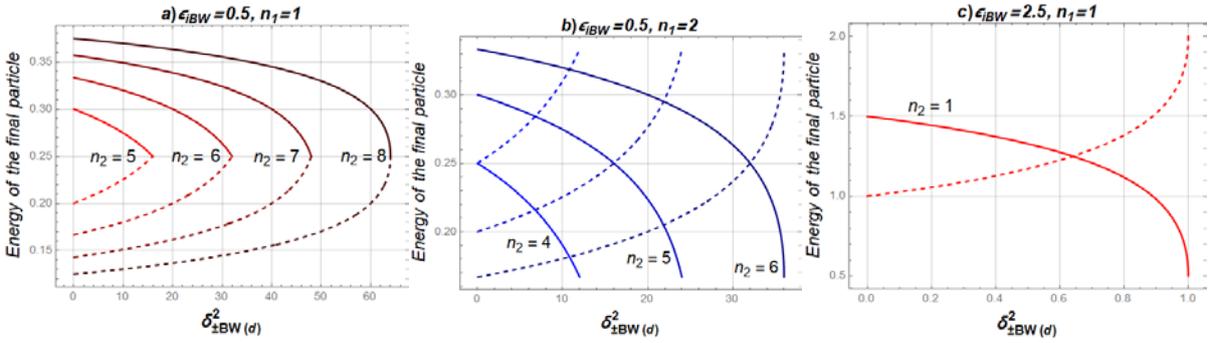

**Figure 4**. Dependence of the energies of the final electron on the upper branch $\varepsilon_{fBW(u)}$ (77) (solid curves) and the electron-positron pair on the lower branch $2\varepsilon_{\pm BW(d)}$ (40) (hatched curves) on the ultra-relativistic parameter of the pair $\delta^2_{\pm BW(d)}$ for a fixed number of the photons of the wave absorbed in the first $(n_1 \geq 1)$ and second $(n_2 \geq \bar{n}_{th})$ vertices and the given energy of the initial electron $\varepsilon_{iBW}$. Fig. 4a and Fig.4b correspond to the cases when the number of photons of the wave at the second vertex begins with a threshold value ($n_2 \geq \bar{n}_{th}$, (79)-(83)). Fig.4c corresponds the big initial electron energies, when $n_2 \geq 1$.

Figure 4 shows the dependence of the energies of the final particles (the electron on the upper branch (77) and the electron-positron pair on the lower branch (40)) on the square of the outgoing angle of the pair for different initial electron energies with fixed number of photons of the wave $n_1$ and $n_2$. Fig. 4a and fig. 4b are plotted for a small initial electron energy $(\varepsilon_{iBW} = 0.5)$, when the number of the photons of the wave absorbed in the second vertex depends on the outgoing angle of the pair and $n_1$ ($n_2 \geq \bar{n}_{th}$, (79)-(83)). At $n_1 = 1$ we have $n_2 = 5,6,7,8,\ldots$ and at $n_1 = 2$ we have $n_2 = 4,5,6,\ldots$. Fig.4c is drawn for big initial electron energy $(\varepsilon_{iBW} = 2.5)$ when the number of the photons of the wave absorbed in the second vertex starts with one. Fig.4a illustrates case when the energy of the electron always exceeds the pair energy. Fig.4b and 4c illustrate cases when the energy of the electron can be higher or lower than the pair energy.

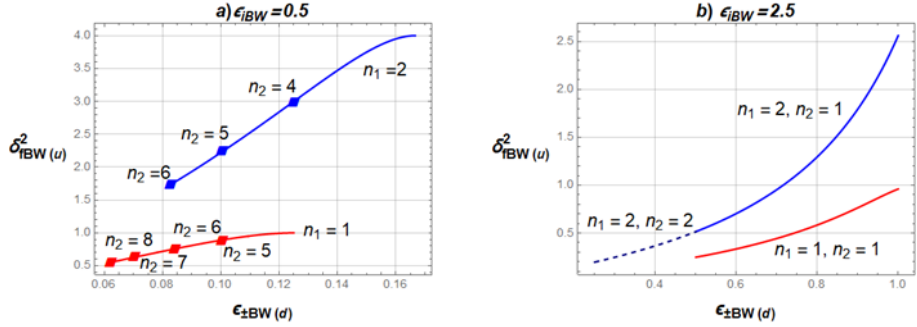

**Figure 5**. Dependence of the ultra-relativistic parameter of the electron $\delta^2_{fBW(u)}$ (72) on the pair energy on the lower branch $\varepsilon_{\pm BW(d)}$ (40) for different numbers of photons of the wave absorbed in the first $(n_1 \geq 1)$ and second $(n_2 \geq \bar{n}_{th})$ vertices (79), (80). Points on fig.5a correspond to the decrease in the threshold number of photons of the wave absorbed in the second vertex with the increase in the pair energy. Figure5b is drawn for large energy of the initial electron when $n_2 \geq 1$.

Figure 5 shows the dependence of the square of the outgoing angle of the electron (72) on the electron-positron pair energy on the lower branch (40). Figure 5a is drawn for small energy of the initial electron $(\varepsilon_{iBW} = 0.5)$, when the threshold number of the photons absorbed in the second vertex depends on the pair energy (the NBWP). Figure5b is drawn for large energy of the initial electron $(\varepsilon_{iBW} = 2.5)$, when the threshold number of the photons absorbed in the second vertex starts with one. Figure 5 shows that each interval of the electron-positron pair energy uniquely corresponds to a certain range of electron outgoing angles.

**4. The resonant trident process rate.**

In the previous section, it was shown that the two final electrons (in the first and second vertices) have different energy under resonance conditions. This is because the electron of the pair has the same energy as the positron. Because of this, the final electron and the electron pair are distinguishable (see Fig.2) and as a result, channels A and B do not interfere and are topologically identical. Therefore, in the future we will consider only channel A. It is important to emphasize that for the resonant channel A the final particle energies for different numbers of photons of the wave are also different (see Fig. 4 and Fig. 5). Therefore, the amplitudes of the process with different numbers of photons of the wave $(n_1, n_2)$ also do not interfere with each other. Given this, in the case of unpolarized initial and final particles, it is not difficult to obtain an expression for the partial differential RTP rate (for detailed calculations, see the article [60]):

$$\frac{dW_{n_1 n_2}}{E_i d\delta_{\pm BW}^2 d\delta_{fC}^2} = \left(\Gamma_C \Gamma_{BW}\right) \frac{16\pi^3 m_*^4}{|q^2|^2 E_i^2} K_{n_1}\left(u_f, z_f\right) P_{n_2}\left(u_\pm, z_\pm\right) \delta\left(E_i - E_f - 2E_\pm\right) E_f dE_f dE_\pm. \quad (90)$$

where $\Gamma_C$ and $\Gamma_{BW}$ are the unitless coefficients that determine the order of magnitude of the NCE and NBWP, when the energy of the initial electrons $E_i \sim \omega_C$:

$$\Gamma_C = \alpha\left(\frac{m}{\omega_C}\right)^2, \quad \Gamma_{BW} = \alpha\left(\frac{m}{\omega_{BW}}\right)^2. \quad (91)$$

In the relation (90) function $K_{n_1}$ defines the probability of the NCE, and the function $P_{n_2}$ defines the probability of the NBWP.

$$K_{n_1}\left(u_f, z_f\right) = -4J_{n_1}^2\left(z_f\right) + \eta^2\left(2 + \frac{u_f^2}{1 + u_f}\right) B_{n_1}\left(z_f\right), \quad (92)$$

$$P_{n_2}\left(z_\pm\right) = J_{n_2}^2\left(z_\pm\right) + \frac{1}{2}\eta^2 B_{n_2}\left(z_\pm\right). \quad (93)$$

In the expressions (92), (93), the function $B_n(z)$ ($n = n_1, z = z_f$; $n = n_2, z = z_\pm$) and the arguments of the Bessel functions $z_f$ and $z_\pm$, as well as the relativistic-invariant parameter $u_f$, have the following form:

$$B_n(z) = J_{n-1}^2(z) + J_{n+1}^2 - 2J_n^2, \quad (94)$$

$$z_f = 2n_1 \frac{\eta}{\sqrt{1+\eta^2}} \sqrt{\zeta_f\left(1 - \zeta_f\right)}, \quad \zeta_f = \frac{1}{n_1 \varepsilon_{iC}}\left(\frac{E_i}{E_f} - 1\right), \quad (95)$$

$$z_\pm = 2n_2 \frac{\eta}{\sqrt{1+\eta^2}} \sqrt{\zeta_\pm\left(1 - \zeta_\pm\right)}, \quad \zeta_\pm = \frac{\omega_{BW}}{2n_2 E_\pm}, \quad (96)$$

$$u_f = \left(\frac{E_i}{E_f} - 1\right). \quad (97)$$

The delta function in the relation (90) makes it easy to integrate over the energy of the final electron or one of the particles of the pair. At the same time, two possible variants of the final particle energies should be taken into account. When the energy of the final electron is on the lower branch $\varepsilon_{fC(d)}$, in the relation (90), integration should be performed over the energy of the particle of the pair. If the energy of the electron (positron) of the pair is on the lower branch $\varepsilon_{\pm BW(d)}$, then in the relation (90), integration should be performed over the energy of the final electron.

*4.1 The differential and complete resonant trident process rate: the lower branch of the electron energy, the upper branch of the pair energy.*

Let us perform integration in relation (90) over the pair particle energy. As a result, we obtain the following expression for the partial differential RTP rate on the lower energy branch of the final electron:

$$\frac{dW_{n_1 n_2}^{(f)}}{dE_{f(d)}} = \left(\Gamma_C \Gamma_{BW}\right) \frac{64\pi^3 m_*^4 \varepsilon_{fC(d)}}{|q^2|^2 \varepsilon_{iC}} K_{n_1}\left(u_f, z_f\right) P_{n_2}\left(u_\pm, z_\pm\right) d\delta_{fC}^2 d\delta_{\pm C}^2, \quad (98)$$

where the functions $K_{n_1}$ and $P_{n_2}$ are defined by the expressions (92) and (93), in which the functions $u_f = u_{fC(d)}$, $z_f = z_{fC(d)}$ are taken on the lower branch of the final electron energy and the function $z_\pm = z_{\pm C(u)}$ is defined on the upper branch of the pair energy (see section 3):

$$u_{fC(d)} = \left(\frac{\varepsilon_{iC}}{\varepsilon_{fC(d)}} - 1\right), \quad z_{\pm C(u)} = \frac{2\eta}{\sqrt{1+\eta^2}}\sqrt{\delta^2_{\pm C(u)}}, \tag{99}$$

$$z_{fC(d)} = 2n_1 \frac{\eta}{\sqrt{1+\eta^2}} \sqrt{\zeta_{fC(d)}(1-\zeta_{fC(d)})}, \quad \zeta_{fC(d)} = \frac{1}{n_1}\left(\frac{1}{\varepsilon_{fC(d)}} - \frac{1}{\varepsilon_{iC}}\right). \tag{100}$$

Here, the function $\varepsilon_{fC(d)}$ changes in the interval (34), and the ultra-relativistic parameter of the pair $\delta^2_{\pm C(u)}$ is determined by the final electron energy on the lower branch (see the relations (51), (52)).

Elimination of the resonant infinity in the expression (98) (98) is performed with the Breit-Wigner procedure [89, 60-64]. Because of this, the zero component and the square of the 4-pulse of the intermediate photon get an imaginary additive:

$$q_0 \to q_0 - i\Gamma_{BW}, \quad q^2 \to q^2 - 2iq_0\Gamma_{BW}, \quad \Gamma_{BW} = \frac{1}{2}W_{BW}. \tag{101}$$

where $W_{BW}$ is the total probability (per unit time) of the NBWP.

$$W_{BW}(\eta, \varepsilon_{iBW}) = \frac{\alpha m^2}{8\pi q_0} P(\eta, \varepsilon_{iBW}), \tag{102}$$

$$P(\eta, \varepsilon_{iBW}) = \sum_{k=k_{\min}=\lceil \varepsilon_{iBW}^{-1} \rceil}^{\infty} P_k(\eta, \varepsilon_{iBW}), \quad P_k(\varepsilon_{iBW}) = \int_1^{k\varepsilon_{iBW}} \frac{du}{u\sqrt{u(u-1)}} P(u, \eta, k\varepsilon_{iBW}), \tag{103}$$

$$P(\eta, u, k\varepsilon_{iBW}) = J_k^2(z) + \eta^2 \left(u - \frac{1}{2}\right)\left[J_{k+1}^2(z) + J_{k-1}^2(z) - 2J_k^2(z)\right], \tag{104}$$

$$z = 2k \frac{\eta}{\sqrt{1+\eta^2}} \sqrt{\frac{u}{k\varepsilon_{iBW}}\left(1 - \frac{u}{k\varepsilon_{iBW}}\right)}. \tag{105}$$

Given this, the differential partial RTP rate (98) takes the following form:

$$\frac{dW_{n_1n_2}^{(f)}}{dE_{f(d)}} = (\Gamma_C \Gamma_{BW}) \frac{64\pi^3}{\varepsilon_{iC}^3 \varepsilon_{fC(d)}} K_{n_1}\left(u_{fC(d)}, z_{fC(d)}\right) P_{n_2}\left(z_{\pm C(u)}\right) \int \frac{d\delta^2_{fC}}{\left[\left(\delta^2_{fC} - \delta^2_{fC(d)}\right)^2 + \Upsilon^2_{f(d)}\right]} d\delta^2_{\pm C}. \tag{106}$$

Here, the angular width of the resonance is determined by the following expression:

$$\Upsilon_{f(d)} = \frac{\alpha m^2 P(\eta, \varepsilon_{iBW})}{8\pi m_*^2 \varepsilon_{iC} \varepsilon_{fC(d)}}. \tag{107}$$

In the relation (106), the ultra-relativistic electron parameter $\delta^2_{fC(d)}$ determines the resonant energy of the final electron (see expressions (32)), and the parameter $\delta^2_{fC}$ can take arbitrary values that do not depend on the energy of the electron. Note that the expression (106) has the characteristic Breit-Wigner resonance structure and takes the maximum value at $\delta_{fC}^2 = \delta_{fC(d)}^2$. Let us integrate the expression (106) over the ultra-relativistic parameter $\delta_{fC}^2$ near the resonant value $\delta_{fC(d)}^2$. It should be taken into account that the resonant width in this expression has a dominant value only in the resonant denominator (in all other expressions, this width can be ignored). Therefore, this integration actually reduces to an integral only of the resonant denominator:

$$\int_{-\infty}^{\infty} \frac{dy}{\left[y^2 + \Upsilon_{f(d)}^2\right]} = \frac{\pi}{\Upsilon_{f(d)}}, \quad \left(y = \delta_{fC}^2 - \delta_{fC(d)}^2\right). \tag{108}$$

Here, the limits of integration are extended to $\pm\infty$ due to the fast convergence of the integral. Given this, the expression for the differential partial RTP rate (106) after this integration will take the form:

$$\frac{1}{\omega_C} \frac{dW_{n_1 n_2}^{(f)}}{d\varepsilon_{fC(d)}} = \left(\Gamma_C \Gamma_{BW}\right) \frac{16(2\pi)^5 \left(1+\eta^2\right)}{\alpha P(\eta, \varepsilon_{iBW}) \varepsilon_{iC}^2} K_{n_1}\left(u_{fC(d)}, z_{fC(d)}\right) P_{n_2}\left(z_{\pm C(u)}\right) d\delta_{\pm C}^2. \tag{109}$$

We perform integration over the outgoing angle of the electron-positron pair $\delta_{\pm C}^2$ near the resonant value $\delta_{\pm C(u)}^2$. This integration can be performed using the Dirac delta function, which selects the desired resonant outgoing angles of the pair.

$$V_{n_1 n_2}^{(f)}\left(\varepsilon_{fC(d)}\right) = \frac{1}{\omega_C} \frac{dW_{n_1 n_2}^{(f)}}{d\varepsilon_{fC(d)}} =$$

$$= \left(\Gamma_C \Gamma_{BW}\right) \frac{16(2\pi)^5 \left(1+\eta^2\right)}{\alpha P(\eta, \varepsilon_{iBW}) \varepsilon_{iC}^2} K_{n_1}\left(u_{fC(d)}, z_{fC(d)}\right) \int_{\delta_{\pm C(u)}^2 - \Delta}^{\delta_{\pm C(u)}^2 + \Delta} P_{n_2}\left(z_{\pm(u)}\right) 2\pi \, \delta\left(\delta_{\pm C}^2 - \delta_{\pm C(u)}^2\right) d\delta_{\pm C}^2. \tag{110}$$

After integration, the expression for the distribution of the partial rete of the RTP as a function of the final electron energy (110) looks like this:

$$V_{n_1 n_2}^{(f)}\left(\varepsilon_{fC(d)}\right) = \Gamma_C \Gamma_{BW} \left(\frac{16 a_i}{\varepsilon_{iC}^2}\right) \Psi_{n_1 n_2}^{(f)}\left(\varepsilon_{fC(d)}\right). \tag{111}$$

where the function $a_i$ is defined by the initial installation parameters:

$$a_i = \frac{(2\pi)^6 \left(1+\eta^2\right)}{\alpha P(\eta, \varepsilon_{iBW})}, \tag{112}$$

and the function $\Psi_{n_1 n_2}^{(f)}$ determines the energy distribution of the final electron:

$$\Psi_{n_1 n_2}^{(f)}\left(\varepsilon_{fC(d)}\right) = K_{n_1}\left(u_{fC(d)}, z_{fC(d)}\right) P_{n_2}\left(z_{\pm C(u)}\right). \tag{113}$$

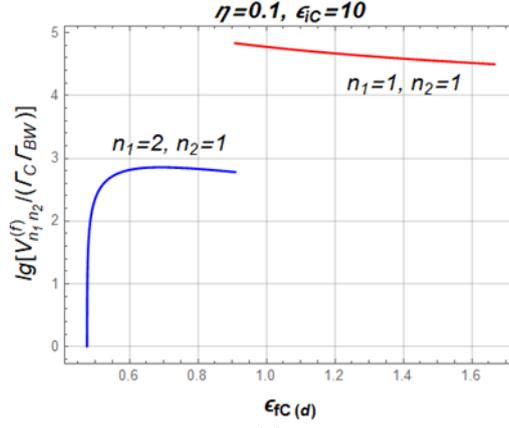

**Figure. 6**. Dependence of the partial RTP rate $V_{n_1 n_2}^{(f)}$ (111) (in units of $\Gamma_C \Gamma_{BW}$) on the energy of the final electron on the lower branch $\varepsilon_{fC(d)}$ for a weak field $(\eta = 0.1)$ and a large energy of the initial electron $\varepsilon_{iC} = 10$ when $n_1 \geq 1, n_2 \geq 1$.

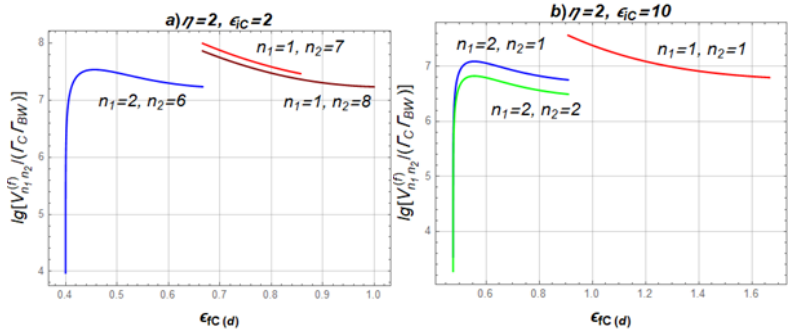

**Figure. 7**. Dependence of the partial RTP rate $V_{n_1 n_2}^{(f)}$ (111) (in units of $\Gamma_C \Gamma_{BW}$) on the energy of the final electron on the lower branch $\varepsilon_{fC(d)}$ for a strong field $(\eta = 2)$ and different numbers of the photons of the wave absorbed in the first $(n_1 \geq 1)$ and second $(n_2 \geq n_{th})$ (54) vertices. Figure 7a corresponds to the energy of the initial electron $\varepsilon_{iC} = 2$, Figure 7b corresponds to the energy of the initial electron $\varepsilon_{iC} = 10$.

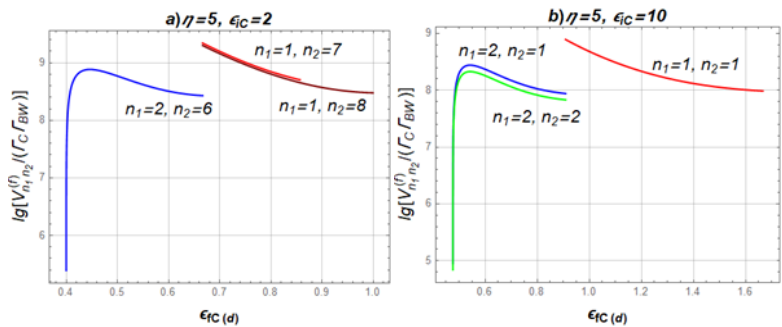

**Figure 8**. Dependence of the partial RTP rate $V_{n_1 n_2}^{(f)}$ (111) (in units of $\Gamma_C \Gamma_{BW}$) on the energy of the final electron on the lower branch $\varepsilon_{fC(d)}$ for a strong field $(\eta = 5)$ and different numbers of the photons of the wave absorbed in the first $\left(n_1 \geq 1\right)$ and second $\left(n_2 \geq n_{th}\right)$ (54) vertices. Figure 7a corresponds to the energy of the initial electron $\varepsilon_{iC} = 2$, Figure 7b corresponds to the energy of the initial electron $\varepsilon_{iC} = 10$.

Figures 6, 7, and 8 show the dependence of the partial RTP rate $V_{n_1 n_2}^{(f)}$ (111) on the energy of the final electron on the lower branch $\varepsilon_{fC(d)}$ (34) for various values of the relativistic-invariant parameter of the wave $\eta$, the initial electron energy, and the number of photons of the wave. Figure 6 shows the case of a weak field $(\eta = 0.1)$ with big initial electron energy $(\varepsilon_{iC} = 10)$ and the number of absorbed photons of the wave $n_1 \geq 1, n_2 \geq 1$. Figure 7 and Figure 8 are plotted for a strong field: $\eta = 2$ and $\eta = 5$, respectively. Figure 7a and Figure 8a show the partial RTP rate $V_{n_1 n_2}^{(f)}$ for small initial electron energy $(\varepsilon_{iC} = 2)$, when the number of photons of the wave absorbed in the second vertex depends on the final electron energy ($n_2 \geq n_{th}$ see the relation (54)). For $n_1 = 1$ we get $n_2 \geq 7$ and for $n_1 = 2$ we get $n_2 \geq 6$. Figure 7b and Figure 8b are plotted for big initial electron energy $(\varepsilon_{iC} = 10)$, when the number of photons of the wave absorbed in the second vertex $n_2 \geq 1$. Figures 6, 7, and 8 show that the partial RTP rate increases significantly as the wave intensity (parameter $\eta$) increases. However, it should also be taken into account that as the parameter $\eta$ increases, the characteristic energy of the NCE also increases, i.e., the unitless coefficient decreases $\Gamma_C \Gamma_{BW}$. Because of this, the effect of the differential RTP rate increase may not appear (see Tables 2 and 4). In addition, for $n_1 = 1$ (regardless of the magnitude $n_2$), the maximum partial RTP rate $V_{n_1 n_2}^{(f)}$ gravitates towards the minimum energy of the final electron (zero outgoing angle $\delta_{fC(d)}^2 = 0$). If $n_1 = 2$ or more, then the maximum partial RTP rate gravitates towards bigger initial electron energies (big outgoing angles $\delta_{fC(d)}^2 > 0$). However, the magnitude of the partial RTP rate rapidly decreases with the increase of the number of absorbed photons of the wave. For those distributions of the partial RTP rate $V_{n_1 n_2}^{(f)}$ on the final electron energy, the pair energy as well as outgoing angles of the electron and the electron-positron pair and uniquely defined by expressions (64), (32) and (51), (52), respectively (also see figures 2 and 3). Thus, we have quantum entanglement of the final particle states, defined by the NCE.

If we integrate relation (111) over the final electron energy, we can obtain the full RTP rate.

$$w^{(f)}(\eta, \varepsilon_{iC}) = \Gamma_C \Gamma_{BW} \left(\frac{16 a_i}{\varepsilon_{iC}^2}\right) \sum_{n_1=1}^{\infty} \int_{\varepsilon_{fC(d)}^{\min}}^{\varepsilon_{fC(d)}^{\max}} d\varepsilon_{fC(d)} K_{n_1}\left(u_{fC(d)}, z_{fC(d)}\right) \sum_{n_2=n_{th}}^{\infty} P_{n_2}\left(z_{\pm C(u)}\right). \quad (114)$$

Here, the limits of electron energy integration vary from the minimum $\left(\varepsilon_{fC(d)}^{\min}\right)$ to the maximum $\left(\varepsilon_{fC(d)}^{\max}\right)$ value (34). We emphasize that when summing up the number of photons of a wave absorbed in the second vertex in the relation (114), it should be taken into account that the threshold value $n_{th}$, generally speaking, depends on the integration variable $\varepsilon_{fC(d)}$. The function argument $P_{n_2}\left(z_{\pm C(u)}\right)$ depends on the electron energy $\varepsilon_{fC(d)}$ and has the form (99), (51), (52). The expression (114) can be made more convenient for calculation by switching to another integration variable: $\varepsilon_{fC(d)} \to \rho_{f(d)}$ (see the relation (36)). After simple calculations, we finally get the following expression for the full RTP rate:

$$w^{(f)}(\eta,\varepsilon_{iC}) = \Gamma_C \Gamma_{BW} \left(\frac{16 a_i}{\varepsilon_{iC}^2}\right) \sum_{n_1=1}^{\infty} \frac{1}{n_1} \int_1^2 d\rho \frac{K_{n_1}\left(u_{fC(\rho)}, z_{fC(\rho)}\right)}{\left(1+\rho/n_1\varepsilon_{iC}\right)^2} \sum_{n_2=n_{th(\rho)}}^{\infty} P_{n_2}\left(z_{\pm C(\rho)}\right). \quad (115)$$

In the expression (115) (115) integration variable $\rho = \rho_{f(d)}$. The arguments of the functions $K_{n_1}\left(u_{fC(\rho)}, z_{fC(\rho)}\right)$ and $P_{n_2}\left(z_{\pm C(\rho)}\right)$, as well as the threshold value $n_{th(\rho)}$ in the expression (115), take the following form (see the relations (99), (100)):

$$u_{fC(\rho)} = \frac{n_1 \varepsilon_{iC}}{\rho}, \quad z_{fC(\rho)} = 2n_1 \frac{\eta}{\sqrt{1+\eta^2}} \sqrt{\frac{1}{\rho}\left(1-\frac{1}{\rho}\right)}, \quad (116)$$

$$z_{\pm C(\rho)} = \frac{2\eta}{\sqrt{1+\eta^2}} \sqrt{\kappa_{C(\rho)}\left(n_2 - \kappa_{C(\rho)}\right)}, \quad (117)$$

$$n_{th(\rho)} = \lceil 2\kappa_{C(\rho)} \rceil, \quad \kappa_{C(\rho)} = \frac{4}{\varepsilon_{iC}}\left(1+\frac{\rho}{n_1\varepsilon_{iC}}\right). \quad (118)$$

It is important to note that when the energy of the initial electron satisfies the condition (61), then the threshold number of photons of the wave absorbed in the second vertex (118) starts with one. In this case, the total RTP rate (115) can be represented as the sum of the partial rates:

$$w^{(f)}(\eta,\varepsilon_{iC}) = \sum_{n_1=1}^{\infty} \sum_{n_2=1}^{\infty} w_{n_1 n_2}^{(f)}, \quad \left(\varepsilon_{iC} \geq \varepsilon_{1(\max)} \approx 9.68\right). \quad (119)$$

where the partial RTP rate with absorption of $n_1$ and $n_2$ photon of the wave has the following form:

$$w_{n_1 n_2}^{(f)} = \left(\Gamma_C \Gamma_{BW}\right)\left(\frac{16 a_i}{\varepsilon_{iC}^2}\right) \frac{1}{n_1} \int_1^2 d\rho \frac{K_{n_1}\left(u_{fC(\rho)}, z_{fC(\rho)}\right)}{\left(1+\rho/n_1\varepsilon_{iC}\right)^2} P_{n_2}\left(z_{\pm C(\rho)}\right). \quad (120)$$

Expressions for partial differential and full rates of the RTP (see relations (111) and (115)) are significantly simplified when the energy of the initial electron significantly exceeds the characteristic energy of the NCE (see the inequality (66)(66)). In this case, the final electron energy is of the same order as the characteristic energy of NCE ($\varepsilon_{fC(d)} \lesssim 1$, see the relation (67)), and the parameter $u_{fC(d)} \sim \varepsilon_{iC} \gg 1$ (see the relation (99)). Because of this, in the function $K_{n_1}\left(u_{fC(d)}, z_{fC(d)}\right)$ (92)(92), the second term will be significantly larger than the first. Therefore, we can write

$$K_{n_1}\left(u_{fC(d)}, z_{fC(d)}\right) \approx \eta^2 \frac{\varepsilon_{iC}}{\varepsilon_{fC(d)}} B_{n_1}\left(z_{fC(d)}\right) \approx \eta^2 \frac{n_1 \varepsilon_{iC}}{\rho_{f(d)}} B_{n_1}\left(z_{fC(d)}\right). \quad (121)$$

Taking this into account, the partial differential RTP rate with registration of the electron energy on the lower branch will take the following form:

$$V^{(f)}_{n_1 n_2}\left(\varepsilon_{fC(d)}\right) \approx \left(\Gamma_C \Gamma_{BW}\right)\left(\eta^2 \frac{16 a_i}{\varepsilon_{iC}}\right) P_{n_2}\left(z_{\pm C(*)}\right) \Psi'^{(f)}_{n_1 n_2}\left(\varepsilon_{fC(d)}\right), \quad (\varepsilon_{iC} \gg 1), \tag{122}$$

$$\Psi'^{(f)}_{n_1 n_2}\left(\varepsilon_{fC(d)}\right) = \frac{1}{\varepsilon_{fC(d)}} B_{n_1}\left(z_{fC(d)}\right). \tag{123}$$

Here the argument of the function $P_{n_2}\left(z_{\pm C(*)}\right)$ does not depend on the energy of the final electron

$$z_{\pm C(*)} = \frac{4\eta}{\sqrt{1+\eta^2}} \sqrt{\frac{n_2}{\varepsilon_{iC}}}, \tag{124}$$

and $B_{n_1}\left(z_{fC(d)}\right)$ has the form:

$$z_{fC(d)} = 2 n_1 \frac{\eta}{\sqrt{1+\eta^2}} \sqrt{\frac{1}{n_1 \varepsilon_{fC(d)}}\left(1 - \frac{1}{n_1 \varepsilon_{fC(d)}}\right)}. \tag{125}$$

Note that the expression (123) (123) contains the electron energy $\varepsilon_{fC(d)} \approx \rho_{f(d)}/n_1 \lesssim 1$. It is important to emphasize that for large initial electron energies, the expression for the partial differential RTP rate (122) decreases approximately hyperbolically with increase of the quantum parameter $\varepsilon_{iC}$.

When condition (66) is fulfilled, we simplify the full RTP rate (115). In this case, the threshold value $n_{th(\rho)} = 1$ (see relation (118)). Therefore, the expression for the total RTP rate can be represented as the sum of the partial RTP rates (119), where the total partial RTP rate with absorption of $n_1$ and $n_2$ photons of the wave after simple transformations will take the following form:

$$w^{(f)}_{n_1 n_2} \approx \left(\Gamma_C \Gamma_{BW}\right)\left(\eta^2 \frac{16 a_i}{\varepsilon_{iC}}\right) P_{n_2}\left(z_{\pm C(*)}\right) \int_1^2 \frac{d\rho}{\rho} B_{n_1}\left(z_{fC(\rho)}\right), \quad (\varepsilon_{iC} \gg 1). \tag{126}$$

In the expression (126), the function argument $B_{n_1}\left(z_{f(\rho)}\right)$ is defined by the relation (116). From the obtained expression (126), it can be seen that the total partial RTP rate with absorption of $n_1$ and $n_2$ photons of the wave decreases with the increase of the initial electron energy as $w^{(f)}_{n_1 n_2} \sim \varepsilon_{iC}^{-1}$.

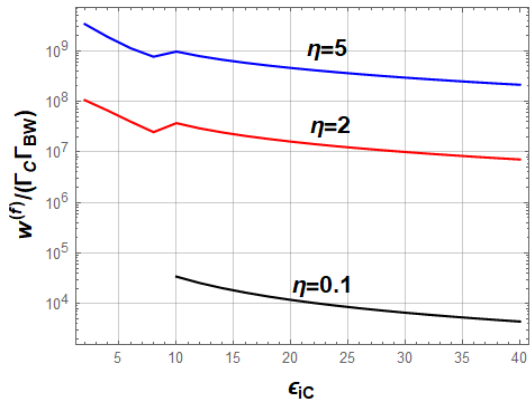

**Figure. 9**. Dependence of the total RTP rate $w^{(f)}$ (115) (in units of $\Gamma_C \Gamma_{BW}$) on the initial electron energy $\varepsilon_{iC}$ for various values of the parameter $\eta$.

Figure 9 shows that the total RTP rate $w^{(f)}$ (115) decreases approximately hyperbolically with the increase of the initial electron energy (in units of the characteristic energy of the NCE). At the same time, as the wave intensity (parameter $\eta$) increases, the total RTP rate rapidly increases. However, it should also be taken into account that as the parameter $\eta$ increases, the characteristic energy of the NCE also increases, i.e., the size coefficient $\Gamma_C \Gamma_{BW}$ decreases. Because of this, the effect of the increase of the total RTP rate $w^{(f)}$ may not be apparent (see Tables 1 and 3). Also note that for the electron energies $2.5 \leq \varepsilon_{iC} \leq 10$, the number of photons of the wave absorbed in the second vertex changes (see the relations (54), (58)-(60)). This effect leads to a characteristic break in the total RTP rate $w^{(f)}$ in this energy interval.

*4.2 The differential and complete resonant trident process rate: the lower branch of the pair energy, the upper branch of the electron energy.*

Let us integrate relation (90) over the final electron energy. As a result, we get the following expression for the partial differential RTP rate:

$$\frac{dW^{(\pm)}_{n_1 n_2}}{dE_{\pm(d)}} = \left(\Gamma_C \Gamma_{BW}\right) \frac{4\pi^3 m_*^4 \left(\varepsilon_{iBW} - 2\varepsilon_{\pm BW(d)}\right)}{\varepsilon_{iBW} \left|q^2\left(\delta^2_{\pm BW}\right)\right|^2} K_{n_1}\left(u_{fBW(u)}, z_{fBW(u)}\right) P_{n_2}\left(z_{\pm BW(d)}\right) d\delta^2_{\pm BW} d\delta^2_{fBW}, \qquad (127)$$

where the functions $K_{n_1}$ and $P_{n_2}$ are defined by the expressions (92) and (93), in which the functions $u_f = u_{fBW(u)}, z_f = z_{fBW(u)}$ are taken on the upper energy branch of the final electron and $z_\pm = z_{\pm BW(d)}$ on the lower energy branch of the pair (see section 3):

$$u_{fBW(u)} = \frac{1}{n_2 \varepsilon_{iBW}/\rho_\tau - 1}, \quad z_{fBW(u)} \approx \frac{\eta}{\sqrt{1+\eta^2}} \sqrt{\delta^2_{fBW(u)}}, \qquad (128)$$

$$z_{\pm BW(d)} = 2n_2 \frac{\eta}{\sqrt{1+\eta^2}} \sqrt{\zeta_{\pm(d)}\left(1-\zeta_{\pm(d)}\right)}, \quad \zeta_{\pm(d)} = \frac{1}{2n_2 \varepsilon_{\pm BW(d)}}. \qquad (129)$$

Here, the energy of the positron (electron) of the pair $\varepsilon_{\pm BW(d)}$ varies in the interval (43), (47), and the ultra-relativistic electron parameter $\delta^2_{fBW(u)}$ is determined by the pair energy on the lower branch (see relations (72), (73)).

Elimination of resonant infinity in the expression (127) is performed with the Breit-Wigner procedure similar to the previous case (see (101) – (105)). After simple conversions, the partial differential RTP (127) rate takes the following form:

$$\frac{dW^{(\pm)}_{n_1 n_2}}{dE_{\pm(d)}} = \left(\Gamma_C \Gamma_{BW}\right) \frac{\pi^3 \left(\varepsilon_{iBW} - 2\varepsilon_{\pm BW(d)}\right)}{\varepsilon_{iBW} \varepsilon^4_{\pm BW(d)}} K_{n_1}\left(u_{fBW(u)}, z_{fBW(u)}\right) P_{n_2}\left(z_{\pm BW(d)}\right) \times$$

$$\times \int \frac{d\delta^2_{\pm BW}}{\left(\delta^2_{\pm BW} - \delta^2_{\pm BW(d)}\right)^2 + \Upsilon^2_{\pm(d)}} d\delta^2_{fBW}. \qquad (130)$$

Here, the angular width of the resonance is determined by the following expression:

$$\Upsilon_{\pm(d)} = \frac{\alpha m^2 P(\eta, \varepsilon_{iBW})}{32\pi m_*^2 \varepsilon_{\pm BW(d)}^2}. \tag{131}$$

In the relation (130), the ultra-relativistic electron parameter $\delta_{\pm BW(d)}^2$ determines the resonant final electron energy (see expressions (32)), and the parameter $\delta_{\pm BW}^2$ can take arbitrary values that don't depend on the energy of the pair. Note that the expression (130) has the characteristic Breit-Wigner resonant structure and takes the maximum value at $\delta_{\pm BW}^2 = \delta_{\pm BW(d)}^2$. Let us integrate the expression (130) over the ultra-relativistic parameters $\delta_{\pm BW}^2$ and $\delta_{fBW}^2$ (see previous section 4.1). After simple calculations, we obtain the following expression for the partial differential RTP rate with registration of the pair energy on the lower branch:

$$V_{n_1 n_2}^{(\pm)}\left(\varepsilon_{\pm BW(d)}\right) = \frac{1}{\omega_{BW}} \frac{dW_{n_1 n_2}^{(\pm)}}{d\varepsilon_{\pm BW(d)}} = \left(\Gamma_C \Gamma_{BW}\right) a_i \Psi_{n_1 n_2}^{(\pm)}\left(\varepsilon_{\pm BW(d)}\right). \tag{132}$$

Here, the function $\Psi_{n_1 n_2}^{(\pm)}$ determines the energy distribution of the electron-positron pair on the lower branch:

$$\Psi_{n_1 n_2}^{(\pm)}\left(\varepsilon_{\pm BW(d)}\right) = \frac{1}{\varepsilon_{\pm BW(d)}^2}\left(1 - \frac{2\varepsilon_{\pm BW(d)}}{\varepsilon_{iBW}}\right) K_{n_1}\left(u_{fBW(u)}, z_{fBW(u)}\right) P_{n_2}\left(z_{\pm BW(d)}\right). \tag{133}$$

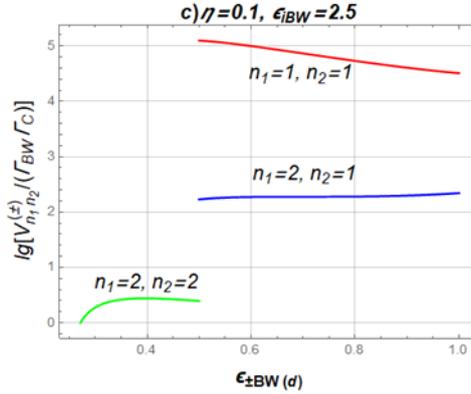

**Figure. 10**. Dependence of the partial differential RTP rate $V_{n_1 n_2}^{(\pm)}$ (132) on the pair energy on the lower branch $\varepsilon_{\pm BW(d)}$ for a weak field $(\eta = 0.1)$ and large initial electron energy $\varepsilon_{iBW} = 2.5$, when $n_1 \geq 1,\, n_2 \geq 1$.

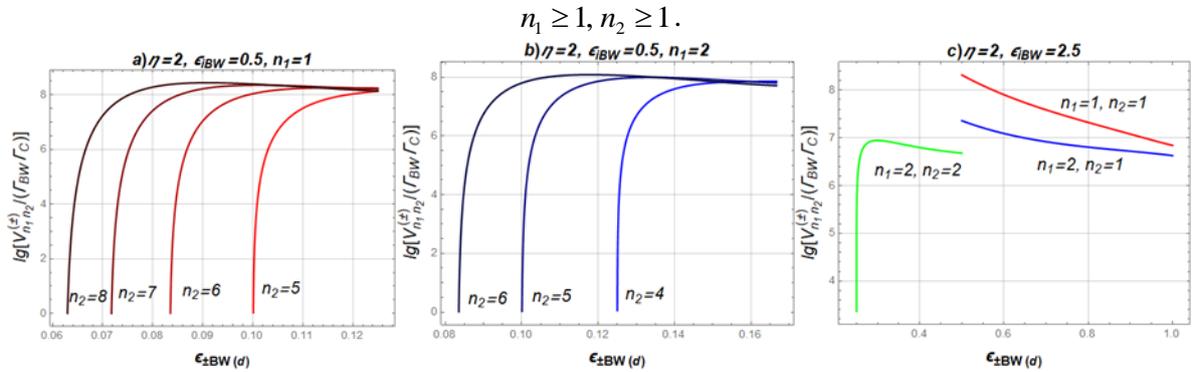

**Figure. 11**. Dependence of the partial differential RTP rate $V_{n_1 n_2}^{(\pm)}$ (132) (in units of $\Gamma_C \Gamma_{BW}$) on the pair energy on the lower branch $\varepsilon_{\pm BW(d)}$ for a strong field $(\eta = 2)$ and different numbers of photons of the wave absorbed in the first $(n_1 \geq 1)$ and second $(n_2 \geq \bar{n}_{th})$ (79), (80) vertices. Fig. 11a and Fig.11b correspond to the initial electron energy $\varepsilon_{iBW} = 0.5$. Figure 11c corresponds to the initial electron energy $\varepsilon_{iBW} = 2.5$.

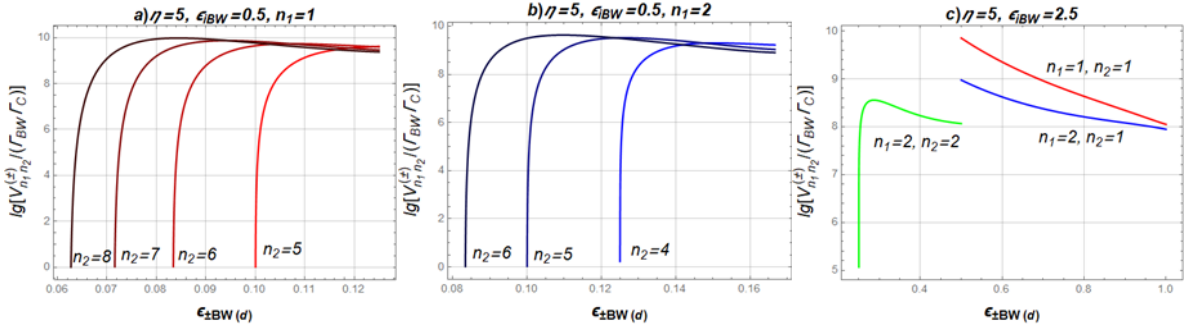

**Figure. 12**. Dependence of the partial differential RTP rate $V_{n_1 n_2}^{(\pm)}$ (in units of $\Gamma_C \Gamma_{BW}$) on the pair energy on the lower branch $\varepsilon_{\pm BW(d)}$ for a strong field $(\eta = 5)$ and different numbers of photons of the wave absorbed in the first $(n_1 \geq 1)$ and second $(n_2 \geq \bar{n}_{th})$ (79), (80) vertices. Fig. 12a and Fig.12b correspond to the initial electron energy $\varepsilon_{iBW} = 0.5$. Figure 12c corresponds to the initial electron energy $\varepsilon_{iBW} = 2.5$.

Figures 10, 11,and 12 show the dependence of the partial differential RTP rate $V_{n_1 n_2}^{(\pm)}$ (132) (in units of $\Gamma_C \Gamma_{BW}$) on the pair energy on the lower branch $\varepsilon_{\pm BW(d)}$ (40), (43) for various values of the relativistic-invariant wave parameter $\eta$, the initial electron energy and the number of absorbed photons of the wave. Figure 10 shows the case of a weak field $(\eta = 0.1)$, big initial electron energy $(\varepsilon_{iBW} = 2.5)$ and the number of absorbed photons of the wave $n_1 \geq 1, n_2 \geq 1$. Fig. 11 and fig. 12 are plotted for strong fields: $\eta = 2$ and $\eta = 5$, respectively. Figures 11a, 11b and figures 12a, 12b show partial differential RTP rate $V_{n_1 n_2}^{(\pm)}$ for small initial electron energy $(\varepsilon_{iBW} = 0.5)$, when the number of photons of the wave absorbed in the second vertex depends on the final electron energy ($n_2 \geq \bar{n}_{th}$, see relations (79), (80)). For $n_1 = 1$ we get $n_2 \geq 5$ and for $n_1 = 2$ we get $n_2 \geq 4$. Fig. 11c and 12c are plotted for big initial electron energies $(\varepsilon_{iBW} = 2.5)$, when the number of photons of the wave absorbed in the second vertex $n_2 \geq 1$. Figures 10, 11, and 12 show that the increase of the parameter $\eta$ by a few orders causes a significant increase in the partial differential RTP rate (by several orders of magnitude). However, it should also be taken into account that as parameter $\eta$ increases the characteristic energy of the NBWP also grows, i.e., the unitless coefficient $\Gamma_C \Gamma_{BW}$ decreases. Thus, the effect of the differential RTP rate increase may not appear (see tables 2 and 4). In addition, for $n_2 = 1$ the maximum rate $V_{n_1 n_2}^{(\pm)}$ holds for the minimum pair energy (zero outgoing angle $\delta_{\pm BW(d)}^2 = 0$). If $n_2 = 2$ or more, the maximum partial

differential RTP rate gravitates towards higher pair energies (bigger outgoing angles $\delta^2_{\pm BW(d)} > 0$). However, the partial RTP rate significantly decreases with the increase of the number of the absorbed photons of the wave. For these energy distributions of the partial differential RTP rate $V^{(\pm)}_{n_1 n_2}$ the final electron energy, outgoing angles of the electron and electron-positron pair are uniquely determined by the relations (77), (78), (72) and (40), (41), respectively (also see fig. 4 and 5). Thus, we have quantum entanglement of the final particle states, determined by the NBWP. If we integrate the relation (132) (111) over all possible pair particle energies (43), we can get the full RTP rate.

$$w^{(\pm)}(\eta, \varepsilon_{iBW}) = a_i (\Gamma_C \Gamma_{BW}) \sum_{n_1=1}^{\infty} \int_{\varepsilon_{\pm BW(d)}^{\min}}^{\varepsilon_{\pm BW(d)}^{\max}} d\varepsilon_{\pm BW(d)} \sum_{n_2=\bar{n}_{th}}^{\infty} \frac{P_{n_2}(z_{\pm BW(d)})}{\varepsilon_{\pm BW(d)}^2} \left(1 - \frac{2\varepsilon_{\pm BW(d)}}{\varepsilon_{iBW}}\right) K_{n_1}(u_{fWB}, z_{fBW}). \quad (134)$$

The expression (134) can be made more convenient for calculation by switching to another integration variable $x_{\pm(d)} \to \rho_{\pm(d)}$ (see the relation (40)). After simple calculations, we finally get the following expression for the full RTP rate:

$$w^{(\pm)}(\eta, \varepsilon_{iBW}) = 2a_i (\Gamma_C \Gamma_{BW}) \sum_{n_1=1}^{\infty} \int_1^2 \frac{d\rho}{\rho^2} \sum_{n_2=\bar{n}_{th}(\rho)}^{\infty} n_2 \left(1 - \frac{\rho}{n_2 \varepsilon_{iBW}}\right) P_{n_2}(z_{\pm BW(\rho)}) K_{n_1}(u_{fWB(\rho)}, z_{fBW(\rho)}). \quad (135)$$

In the expression (135) (135) integration variable $\rho = \rho_{\pm(d)}$. The arguments of the functions $P_{n_2}(z_{\pm BW(\rho)})$ and $K_{n_1}(u_{fBW(\rho)}, z_{fBW(\rho)})$, as well as the threshold value $\bar{n}_{th}$ take the following form (see the relations (95) – (97)):

$$z_{\pm BW(\rho)} = 2n_2 \frac{\eta}{\sqrt{1+\eta^2}} \sqrt{\frac{1}{\rho}\left(1 - \frac{1}{\rho}\right)}, \quad (136)$$

$$z_{fBW(\rho)} = 2n_1 \frac{\eta}{\sqrt{1+\eta^2}} \sqrt{\frac{u_{fBW(\rho)}}{n_1 \varepsilon_{iC}}\left(1 - \frac{u_{fBW(\rho)}}{n_1 \varepsilon_{iC}}\right)}, \quad u_{fBW(\rho)} = \left(\frac{n_2 \varepsilon_{iBW}}{\rho} - 1\right)^{-1}, \quad (137)$$

$$\bar{n}_{th(\rho)} = \lceil \zeta_{\pm(\rho)} \rceil, \quad \zeta_{\pm(\rho)} = \frac{\rho}{\varepsilon_{iBW}}\left(1 + \frac{2}{n_1 \varepsilon_{iC}}\right). \quad (138)$$

It is important to note that if the quantum parameter of the NBWP satisfies the condition (61), then the threshold number of photons of the wave absorbed in the second vertex (138) starts with one. In this case, the full RTP rate can be represented as a sum of partial rates:

$$w^{(\pm)}(\eta, \varepsilon_{iBW}) = \sum_{n_1=1}^{\infty} \sum_{n_2=1}^{\infty} w^{(\pm)}_{n_1 n_2}, \quad \left(\varepsilon_{iBW} \geq \varepsilon_{1(\max)} \approx 2.42\right), \quad (139)$$

where the partial RTP rate with absorption of $n_1$ and $n_2$ photon of the wave has the following form:

$$w^{(\pm)}_{n_1 n_2} = 2a_i (\Gamma_C \Gamma_{BW}) n_2 \int_1^2 \frac{d\rho}{\rho^2}\left(1 - \frac{\rho}{n_2 \varepsilon_{iBW}}\right) P_{n_2}(z_{\pm BW(\rho)}) K_{n_1}(u_{fBW(\rho)}, z_{fBW(\rho)}). \quad (140)$$

Expressions for the differential and full RTP rates (see the relations (132) and (139), (140)) are significantly simplified when the initial electron energy significantly exceeds the characteristic NBWP energy (see inequality (86), (86)). In this case, the energy of the positron (electron) of the pair is small ($\varepsilon_{\pm BW(d)} \lesssim 1$ see (87)). The expressions (137), (138) take the following form:

$$u_{fBW(\rho)} \approx \frac{\rho}{n_2 \varepsilon_{iBW}} \ll 1, \quad z_{fBW(\rho)} \approx \left(\frac{\eta}{\sqrt{1+\eta^2}} \sqrt{\frac{n_1}{n_2}} \rho\right) \frac{1}{\varepsilon_{iBW}} \ll 1. \quad (141)$$

Because of this, the function $K_{n_1}\left(u_{fBW(\rho)}, z_{fBW(\rho)}\right)$ (92) (92) takes the main value when $n_1 = 1$. Therefore, we can write

$$K_{n_1}\left(u_{fBW(\rho)}, z_{fBW(\rho)}\right) \approx K_1\left(u_{fBW(\rho)}, z_{fBW(\rho)}\right) \approx 2\eta^2. \quad (142)$$

Taking this into account, the partial differential RTP rate with registration of the positron (electron) energy of the pair on the lower branch (132) will take the following form:

$$V_{n_1 n_2}^{(\pm)}\left(\varepsilon_{\pm BW(d)}\right) \approx \left(\Gamma_C \Gamma_{BW}\right) 2\eta^2 a_i \Psi'^{(\pm)}_{1n_2}\left(\varepsilon_{\pm BW(d)}\right), \quad \left(\varepsilon_{iBW} \gg 1\right), \quad (143)$$

$$\Psi'^{(\pm)}_{1n_2}\left(\varepsilon_{\pm BW(d)}\right) = \frac{P_{n_2}\left(z_{\pm BW(d)}\right)}{\varepsilon^2_{\pm BW(d)}}. \quad (144)$$

Note that the argument of the function $P_{n_2}\left(z_{\pm BW(d)}\right)$ takes the form. Thus, the expression for the partial differential RTP rate (143) barely depends on the quantum parameter $\varepsilon_{iBW}$.

Let us simplify the full RTP rate (135) when the condition (86) is met. In this case, the threshold value $\bar{n}_{th} \approx \lceil \rho_\tau / \varepsilon_{iBW} \rceil = 1$ (see relation (80)). In addition, the main contribution to the sum over the number of photons of the wave absorbed in the first vertex is given by the term with $n_1 = 1$. Because of this, the relation for the total resonant rate (135) can be roughly represented as the sum of only the number of photons of the wave absorbed in the second vertex:

$$w^{(\pm)}\left(\eta, \varepsilon_{iBW}\right) = \sum_{n_2=1}^{\infty} w^{(\pm)}_{1n_2}, \quad \left(\varepsilon_{iBW} \gg 1\right). \quad (145)$$

where the total partial RTP rate with absorption of $n_2$ - photons of the wave after simple transformations will take the following form:

$$w^{(\pm)}_{1n_2} = \left(\Gamma_C \Gamma_{BW}\right) 4\eta^2 a_i n_2 \int_1^2 \frac{d\rho}{\rho^2} P_{n_2}\left(z_{\pm BW(\rho)}\right). \quad (146)$$

In the expression (146), the function argument $P_{n_2}\left(z_{\pm BW(d)}\right)$ is defined by the relation (136).

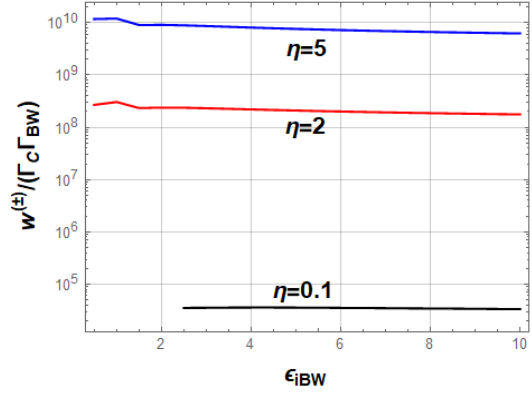

**Figure. 13**. Dependence of the total RTP rate $w^{(\pm)}$ (135) on the initial electron energy $\varepsilon_{iBW}$ for various parameter $\eta$ values.

Figure 13 shows that with the increase of the initial electron energy (in units of characteristic energy of the NBWP) the full RTP rate $w^{(\pm)}$ (135) barely changes. At the same time, with the increase of wave intensity (parameter $\eta$) the full RTP rate increases rapidly. However, it should also be taken into account that as the parameter $\eta$ increases, the characteristic energy of the NBWP also increases, i.e., the unitless coefficient $\Gamma_C \Gamma_{BW}$ decreases. Because of this, the effect of the full RTP rate $w^{(\pm)}$ increase may not occur (see tables 2 and 4). Note that for the initial electron energies $0.5 \leq \varepsilon_{iBW} \leq 2.5$ the number of photons of the wave in the second vertex changes (see relations (79), (81) – (83)). This effect leads to a characteristic break in the total RTP rate $w^{(\pm)}$ in this energy interval.

## 5. Results and discussion.

In the previous sections, resonant kinematics were studied in detail. Distributions of the differential partial RTP rate as a function of the final particle energy and the corresponding total RTP rates as a function of the initial electron energy and parameter $\eta$ were obtained. It is important to note that the obtained resonant relations of the RTP depend on relativistic-invariant parameters $\varepsilon_{jC}(\varepsilon_{jBW})(j=i,f,\pm)$ (the energies of the initial and final particles in units of the characteristic energies of the NCE and NBWP) and the parameter $\eta$.

We estimate the final particle energies and the RTP rates in the optical and X-ray frequencies of the external electromagnetic wave. In this case, we will choose head-on directions of propagation of the initial electrons and the electromagnetic wave $(\theta_i = \pi)$. Note that the choice of small angles $(\theta_i \ll 1)$ when the initial electrons collide with an electromagnetic wave will significantly increase the characteristic energy of the NCE (11), which will lead to suppression of the resonant trident process [56].

### 5.1 Optical frequency range: main parameters of the resonant trident process.

Let us choose the frequency of the electromagnetic wave propagating towards the initial electron flow, $\omega = 3$ eV. Then, depending on the wave intensity $(\eta = 0.1; 2; 5)$, we obtain the following values of the characteristic energy NCE:

$$\omega_C \approx \begin{cases} 21.98 \text{ GeV, if } \omega = 3 \text{ eV}, I = 1.675 \cdot 10^{17} \text{ Wcm}^{-2} (\eta = 0.1) \\ 108.80 \text{ GeV, if } \omega = 3 \text{ eV}, I = 6.70 \cdot 10^{19} \text{ Wcm}^{-2} (\eta = 2) \\ 565.76 \text{ GeV, if } \omega = 3 \text{ eV}, I = 4.19 \cdot 10^{20} \text{ Wcm}^{-2} (\eta = 5) \end{cases} \qquad (147)$$

The relation (147) shows that the characteristic energy $\omega_C$ in the optical frequency range is quite large. Even for a weak field $(\eta = 0.1)$, it has an order of magnitude of tens of GeV, and for $\eta \sim 1$ the characteristic energy $\omega_C \sim 10^2$ GeV. At the same time, the characteristic NBWP energy is four times greater $\omega_C$. As was shown earlier, the main contribution to the RTP rate is made by the initial electron energies $\varepsilon_{iC} \gtrsim 1 (\varepsilon_{iBW} \gtrsim 1)$. Therefore, to observe the resonant trident process in the region of optical frequencies and strong fields, sufficiently large initial electron energies are required $E_i \gtrsim (10^2 \div 10^3)$ GeV.

**Table 1.** The main characteristics of the resonant trident process, defined by the nonlinear Compton effect at $\varepsilon_{iC} = 10$, for the optical frequency range $(\omega = 3 \text{ eV})$ and different wave intensities (147).

| $\omega_C$ (GeV) | $E_i$ (GeV) | $E_{f(d)}^{(min)}$ (GeV) | $E_{f(d)}^{(max)}$ (GeV) | $\Gamma_C \Gamma_{BW}$ | $V_{11(max)}^{(f)}$ | $w^{(f)}$ |
|---|---|---|---|---|---|---|
| 21.98 | 219.8 | | | | | $3.36 \times 10^{-20}$ |
| | | 19.98 | 36.63 | $9.73 \times 10^{-25}$ | $6.61 \times 10^{-20}$ | |
| 108.80 | 1088.0 | | | | | $6.06 \times 10^{-20}$ |
| | | 98.91 | 181.33 | $1.62 \times 10^{-27}$ | $5.93 \times 10^{-20}$ | |
| 565.76 | 5657.6 | | | | | $2.16 \times 10^{-21}$ |
| | | 514.33 | 942.94 | $2.22 \times 10^{-30}$ | $1.74 \times 10^{-21}$ | |

**Table 2.** The main characteristics of the resonant trident process, defined by the nonlinear Breit-Wheeler process at $\varepsilon_{iBW} = 2.5$, for the optical frequency range $(\omega = 3 \text{ eV})$ and different wave intensities (147).

| $\omega_{BW}$ (GeV) | $E_i$ (GeV) | $2E_{\pm(d)}^{(min)}$ (GeV) | $2E_{\pm(d)}^{(max)}$ (GeV) | $\Gamma_C \Gamma_{BW}$ | $V_{11(max)}^{(\pm)}$ | $w^{(\pm)}$ |
|---|---|---|---|---|---|---|
| 87.92 | 219.8 | | | | | $3.44 \times 10^{-20}$ |
| | | 87.91 | 175.8 | $9.73 \times 10^{-25}$ | $1.22 \times 10^{-19}$ | |
| 435.20 | 1088.0 | | | | | $3.84 \times 10^{-19}$ |
| | | 435.2 | 870.4 | $1.62 \times 10^{-27}$ | $3.36 \times 10^{-19}$ | |
| 2263.04 | 5657.6 | | | | | $1.97 \times 10^{-20}$ |
| | | 2263.1 | 4526.1 | $2.22 \times 10^{-30}$ | $1.59 \times 10^{-20}$ | |

Tables 1 and 2 show the main characteristics of the resonant trident process for the optical frequency range $(\omega = 3 \text{ eV})$, different intensities of the wave (147) and high initial electron energy $(\varepsilon_{iC} = 10, \varepsilon_{iBW} = 2.5)$. These tables reflect two possible cases of final particle energies (see Sections 3.2 and 3.3).

It is important to note that for a weak field $(\eta \ll 1)$, there is a threshold energy $\varepsilon_{iC} \geq \varepsilon_{n_1(\min)} = 4(1+\sqrt{2})$ (61), below which the RTP is suppressed (see Figures 6 and 10). Because of this, the energies of the initial electrons must be sufficiently large in comparison with the characteristic energy of the NCE. So, for the energy of the initial electrons $E_i = 219.8 \text{ GeV}$ the energies of the final particles are distributed in the following intervals: $19.98 \text{ GeV} \leq E_{f(d)} \leq 36.63 \text{ GeV}$ and $183.17 \text{ GeV} \leq 2E_{\pm(u)} \leq 199.82 \text{ GeV}$. In this case, since the quantum parameter $\varepsilon_{iC}$ is large, the initial electron energy mainly passes into the electron-positron pair energy with differential and total RTP rates $V_{11}^{(f)} \lesssim 6.61 \times 10^{-20}$ and $w^{(f)} \leq 3.36 \times 10^{-20}$, respectively. In the second case (see Table 2 and Figure 10), the initial electron energy is redistributed between the final particles as follows: $87.91 \text{ GeV} \leq 2E_{\pm(d)} \leq 175.80 \text{ GeV}$ and $44.00 \text{ GeV} \leq E_{f(u)} \leq 131.89 \text{ GeV}$ with the differential and total RTP rates $V_{11}^{(\pm)} \lesssim 1.22 \times 10^{-19}$ and $w^{(\pm)} \leq 3.44 \times 10^{-20}$, respectively.

For a strong field $(\eta = 2)$ and small initial electron energies $(\varepsilon_{iC} = 2)$, the initial electron energy $E_i = 217.6 \text{ GeV}$ is distributed among the final particles as follows. In the first case, we have $72.5 \text{ GeV} \leq E_{f(d)} \leq 108.8 \text{ GeV}$ and $108.8 \text{ GeV} \leq 2E_{\pm(u)} \leq 145.1 \text{ GeV}$ with the differential and total RTP rates $V_{17}^{(f)} \leq 7.08 \times 10^{-20}$ and $w^{(f)} \leq 1.73 \times 10^{-19}$, respectively (see Fig. 7a and Fig. 9). In the second case, we have $108.8 \text{ GeV} \leq 2E_{\pm(d)} \leq 163.2 \text{ GeV}$ and $54.4 \text{ GeV} \leq E_{f(u)} \leq 108.8 \text{ GeV}$ with the differential and total RTP rates $V_{18}^{(\pm)} \leq 4.45 \times 10^{-19}$ and $w^{(\pm)} \leq 4.36 \times 10^{-19}$, respectively (see Fig. 11a and Fig. 13). For large initial electron energy $E_i = 1.088 \text{ TeV}$ $(\varepsilon_{iC} = 10)$ the final particle energy mainly goes into the electron-positron pair energy (see relations (66), (67)). In the first case, we have $98.91 \text{ GeV} \leq E_{f(d)} \leq 181.33 \text{ GeV}$ and $906.67 \text{ GeV} \leq 2E_{\pm(u)} \leq 989.09 \text{ GeV}$ with the differential and total RTP rates $V_{11}^{(f)} \lesssim 5.93 \times 10^{-20}$ and $w^{(f)} \leq 6.06 \times 10^{-20}$, respectively (see table 1, Figures 7b and 9). In the second case, we have $432.20 \text{ GeV} \leq 2E_{\pm(d)} \leq 870.40 \text{ GeV}$ and $217.60 \text{ GeV} \leq E_{f(u)} \leq 652.80 \text{ GeV}$ with the differential and total RTP rates $V_{11}^{(\pm)} \lesssim 3.36 \times 10^{-19}$ and $w^{(\pm)} \leq 3.84 \times 10^{-19}$, respectively (see table 2, Figures 11c).

It follows that in the region of optical frequencies of the electromagnetic wave, for the initial electron energies $E_i \gtrsim \omega_C \gtrsim (10 \div 10^2) \text{ GeV}$ the differential and total RTP rates have the following order of magnitude: $V_{n_1 n_2}^{(f)} \lesssim V_{n_1 n_2}^{(\pm)} \lesssim 10^{-19}$ and $w^{(f)} \lesssim w^{(\pm)} \lesssim 10^{-19}$, respectively. Note that in the case of a non-resonant trident process, the corresponding differential rates, which was defined as the product of the differential rates of the NCE and NBWP at $\eta = 3$, $E_i = 17.5\,\text{GeV}$ is of the order of magnitude $\lesssim 10^{-25}$ [1, 5, 56] (under these initial conditions, the resonant trident process is suppressed).

*5.2 X-ray frequency range: main parameters of the resonant trident process.*

As the wave frequency increases, the characteristic energy of the NCE decreases (11), which makes for more favorable conditions for experimental observation of the resonant trident process. Let us choose the frequency of the electromagnetic wave propagating towards the initial electron flow, $\omega = 3\,\text{keV}$. Then, depending on the wave intensity $(\eta = 0.1; 2; 5)$, we obtain the following values of the characteristic energy NCE:

$$\omega_C \approx \begin{cases} 21.98\,\text{MeV, if }\ \omega = 3\,\text{keV},\ I = 1.675 \cdot 10^{23}\,\text{Wcm}^{-2}\ (\eta = 0.1) \\ 108.80\,\text{MeV, if }\ \omega = 3\,\text{keV},\ I = 6.70 \cdot 10^{25}\,\text{Wcm}^{-2}\ (\eta = 2) \\ 565.76\,\text{MeV, if }\ \omega = 3\,\text{keV},\ I = 4.19 \cdot 10^{26}\,\text{Wcm}^{-2}\ (\eta = 5) \end{cases} \quad (148)$$

From relation (148) it can be seen that the characteristic energy $\omega_C$ in the X-ray frequency range is three orders of magnitude smaller than the corresponding values for the optical frequency range. So, for a weak field $(\eta = 0.1)$ it has the order of magnitude of tens of MeV, and for $\eta \sim 1$ the characteristic energy is $\omega_C \sim 10^2\,\text{MeV}$. Therefore, to observe the resonant trident process in the region of X-ray frequencies and strong fields, not very large energies of the initial electrons are required $E_i \gtrsim (10^2 \div 10^3)\,\text{MeV}$.

**Table 3.** Main characteristics of the resonant trident process, determined by the nonlinear Compton effect at $\varepsilon_{iC} = 10$, for the X-ray frequency range $(\omega = 3\,\text{keV})$ and different wave intensities (148).

| $\omega_C$ (MeV) | $E_i$ (MeV) | $E_{f(d)}^{(min)}$ (MeV) | $E_{f(d)}^{(max)}$ (MeV) | $\Gamma_C \Gamma_{BW}$ | $V_{11(max)}^{(f)}$ | $w^{(f)}$ |
|---|---|---|---|---|---|---|
| 21.98 | 219.8 | | | | | $3.36 \times 10^{-8}$ |
| | | 19.98 | 36.63 | $9.732 \times 10^{-13}$ | $6.61 \times 10^{-8}$ | |
| 108.80 | 1088.0 | | | | | $6.06 \times 10^{-8}$ |
| | | 98.91 | 181.33 | $1.620 \times 10^{-15}$ | $5.93 \times 10^{-8}$ | |
| 565.76 | 5657.6 | | | | | $2.16 \times 10^{-9}$ |
| | | 514.33 | 942.94 | $2.216 \times 10^{-18}$ | $1.74 \times 10^{-9}$ | |

**Table 4.** Main characteristics of the resonant trident process, determined by the nonlinear Breit-Wheeler process at $\varepsilon_{iBW} = 2.5$, for the X-ray frequency range $(\omega = 3 \text{ keV})$ and different wave intensities (148).

| $\omega_{BW}$ (MeV) | $E_i$ (MeV) | $2E_{\pm(d)}^{(\min)}$ (MeV) | $2E_{\pm(d)}^{(\max)}$ (MeV) | $\Gamma_C \Gamma_{BW}$ | $V_{11(\max)}^{(\pm)}$ | $w^{(\pm)}$ |
|---|---|---|---|---|---|---|
| 87.92 | 219.8 | | | | | $3.44 \times 10^{-8}$ |
| | | 87.91 | 175.8 | $9.732 \times 10^{-13}$ | $1.22 \times 10^{-7}$ | |
| 435.20 | 1088.0 | | | | | $3.84 \times 10^{-7}$ |
| | | 435.2 | 870.4 | $1.620 \times 10^{-15}$ | $3.36 \times 10^{-7}$ | |
| 2263.04 | 5657.6 | | | | | $1.97 \times 10^{-8}$ |
| | | 2263.1 | 4526.1 | $2.216 \times 10^{-18}$ | $1.59 \times 10^{-8}$ | |

Tables 3 and 4 show the main characteristics of the resonant trident process for the X-ray frequency range $(\omega = 3 \text{ keV})$, different wave intensities (147) and high initial electron energy $(\varepsilon_{iC} = 10, \varepsilon_{iBW} = 2.5)$. These tables reflect two possible cases of final particle energies (see Sections 3.2 and 3.3).

For a weak X-ray wave field $(\eta = 0.1)$ at $\varepsilon_{iC} = 10$ the initial electron energy is $E_i = 219.8 \text{ MeV}$. Then in the first case (see table 3 and figure 6) the final particle energies are distributed in the following range: $19.98 \text{ MeV} \leq E_{f(d)} \leq 36.63 \text{ MeV}$ and $183.17 \text{ MeV} \leq 2E_{\pm(u)} \leq 199.82 \text{ MeV}$. In this case, since the quantum parameter $\varepsilon_{iC}$ is large, the initial electron energy mainly goes into the electron-positron pair energy with the differential and total RTP rates $V_{11}^{(f)} \lesssim 6.61 \times 10^{-8}$ and $w^{(f)} \leq 3.36 \times 10^{-8}$, respectively. In the second case (see table 4 and figure 10) the initial electron energy is redistributed between the final particles as follows: $87.91 \text{ MeV} \leq 2E_{\pm(d)} \leq 175.80 \text{ MeV}$ and $44.00 \text{ MeV} \leq E_{f(u)} \leq 131.89 \text{ MeV}$ with the differential and total RTP rates $V_{11}^{(\pm)} \lesssim 1.22 \times 10^{-7}$ and $w^{(\pm)} \leq 3.44 \times 10^{-8}$, respectively.

For a strong field $(\eta = 2)$ and small quantum parameter $(\varepsilon_{iC} = 2)$, the initial electron energy $E_i = 217.6 \text{ MeV}$ is distributed between final particles the following way. In the first case, we have $72.5 \text{ MeV} \leq E_{f(d)} \leq 108.8 \text{ MeV}$ and $108.8 \text{ MeV} \leq 2E_{\pm(u)} \leq 145.1 \text{ MeV}$ with the differential and total RTP rates $V_{17}^{(f)} \leq 7.08 \times 10^{-8}$ and $w^{(f)} \leq 1.73 \times 10^{-7}$, respectively (see figure 7a and figure 9). In the second case, we have $108.8 \text{ MeV} \leq 2E_{\pm(d)} \leq 163.2 \text{ MeV}$ and $54.4 \text{ MeV} \leq E_{f(u)} \leq 108.8 \text{ MeV}$ with the differential and total RTP rates $V_{18}^{(\pm)} \leq 4.45 \times 10^{-7}$ and $w^{(\pm)} \leq 4.36 \times 10^{-7}$, respectively (see figure 11a and figure 13). For the five times bigger initial electron energy $E_i = 1.088 \text{ GeV}$ $(\varepsilon_{iC} = 10)$ the initial electron energy mainly goes into the electron-positron pair energy (see relations (66), (67)). In the first case, we have $98.91 \text{ MeV} \leq E_{f(d)} \leq 181.33 \text{ MeV}$ and $906.67 \text{ MeV} \leq 2E_{\pm(u)} \leq 989.09 \text{ MeV}$ with the differential and total RTP rates $V_{11}^{(f)} \lesssim 5.93 \times 10^{-8}$ and $w^{(f)} \leq 6.06 \times 10^{-8}$, respectively (see table 1, figures 7b and 9). In the second case, we have $432.20 \text{ MeV} \leq 2E_{\pm(d)} \leq 870.40 \text{ MeV}$ and $217.60\,\text{MeV} \leq E_{f(u)} \leq 652.80\,\text{MeV}$ with the differential and total RTP rates $V_{11}^{(\pm)} \lesssim 3.36 \times 10^{-7}$ and $w^{(\pm)} \lesssim 3.84 \times 10^{-7}$, respectively (see table 4 and figure 11c).

It follows that in the region of X-ray frequencies of an electromagnetic wave, for the initial electron energies $E_i \gtrsim \omega_C \gtrsim (10 \div 10^2)\,\text{MeV}$ the differential and total RTP rates have the following order of magnitude: $V_{n_1 n_2}^{(f)} \lesssim V_{n_1 n_2}^{(\pm)} \lesssim 10^{-7}$ и $w^{(f)} \lesssim w^{(\pm)} \lesssim 10^{-7}$, respectively. Thus, for the X-ray frequency range of an electromagnetic wave, there is a very high differential and total resonant trident process rate.

In conclusion, we formulate the main results:

1. The resonant trident process takes place when an intermediate gamma-quantum in the wave field enters the mass envelope. As a result, the initial second-order process effectively splits into two first-order processes according to the fine structure constant: the nonlinear Compton effect and the nonlinear Breit-Wheeler process. At the same time, the intermediate real gamma-quantum is unobservable, in the quantum mechanical sense.

2. The resonant trident process is determined by two characteristic quantum energies: the characteristic energy of the nonlinear Compton effect $\omega_C$ and the characteristic energy of the nonlinear Breit-Wheeler process $\omega_{BW} = 4\omega_C$ (11). These characteristic energies depend significantly on the frequency and intensity of the wave, as well as on the angle between the momenta of the initial electrons and the electromagnetic wave. So, if we direct the flow of initial electrons towards an electromagnetic wave $(\theta_i = \pi)$, then in the field of optical frequencies $(\omega \sim 1\,\text{eV})$ and wave intensities $I \sim 10^{20}\,\text{Wcm}^{-2}\,(\eta \sim 1)$, the characteristic energy of the nonlinear Compton effect is quite large: $\omega_C \sim 10^2\,\text{GeV}$. At the same time, if we switch to the X-ray range of frequencies $(\omega \sim 1\,\text{keV})$ and wave intensities $I \sim 10^{26}\,\text{Wcm}^{-2}\,(\eta \sim 1)$, then the characteristic energy of the nonlinear Compton effect will be three orders of magnitude less $\omega_C \sim 10^2\,\text{MeV}$.

3. The resonant trident process is effective when the energy of the initial electrons is greater than or on the order of magnitude of the corresponding characteristic energies. At the same time, there are two possible cases of resonant kinematics, when quantum entanglement of final particles takes place. In the first case, the outgoing angles of the final particles, as well as the energy of the electron (positron) of the pair, are uniquely determined by the energy of the final electron (the nonlinear Compton effect at the first vertex, the quantum parameter $\varepsilon_{iC}$ (10)). In the second case, the outgoing angles of the final particles, as well as the energy of the final electron, are uniquely determined by the energy of the electron-positron pair (the nonlinear Breit-Wheeler process at the second vertex, a quantum parameter $\varepsilon_{iBW}$). An important aspect of the resonant trident process is the equality of the energies of the electron and positron pairs.

4. Analytical expressions are obtained for the partial differential rates of RTP from the energy of a final electron $V^{(f)}_{n_1 n_2}$ (in the first case of quantum entanglement of particles) and the energy of an electron-positron pair $V^{(\pm)}_{n_1 n_2}$ (in the second case of quantum entanglement of particles). In the field of optical and X-ray frequencies, these distributions have the following order of magnitude $V^{(f)}_{n_1 n_2} \lesssim V^{(\pm)}_{n_1 n_2} \lesssim 10^{-19}$ and $V^{(f)}_{n_1 n_2} \lesssim V^{(\pm)}_{n_1 n_2} \lesssim 10^{-7}$, respectively.

5. Analytical expressions for the total RTP rate are obtained, depending on the classical parameter $\eta$, as well as on the quantum parameter $\varepsilon_{iC}$ (in the first case, the quantum entanglement of particles) and the quantum parameter $\varepsilon_{iBW}$ (in the second case, the quantum entanglement of particles). In the field of optical and X-ray frequencies, these total RTP speeds are of the following order of magnitude $w^{(f)} \lesssim w^{(\pm)} \lesssim 10^{-19}$ and $w^{(f)} \lesssim w^{(\pm)} \lesssim 10^{-7}$, respectively.

The results obtained can be used in experiments at leading laser centers, as well as to explain QED processes in strong X-ray fields near neutron stars and magnetars [1, 2, 4, 5, 66-71].